\documentclass[aps,prb,showpacs,preprintnumbers,twocolumn,superscriptaddress,nofootinbib]{revtex4-2}
\usepackage{amsmath,amssymb}
\usepackage{bm}
\usepackage{tipa}
\usepackage{upgreek}
\usepackage{comment}
\usepackage{mathrsfs}
\usepackage{graphicx}
\usepackage{braket}
\usepackage{enumitem}
\usepackage{mathbbol}
\usepackage{gensymb}
\usepackage[normalem]{ulem}
\usepackage{color}
\usepackage[colorlinks,bookmarks=true,citecolor=blue,linkcolor=red,urlcolor=blue]{hyperref}
\usepackage{hyperref}

\usepackage{pifont}

\usepackage{todonotes}
\newcommand{\be}{\begin{equation}}
\newcommand{\ee}{\end{equation}}
\newcommand{\bea}{\begin{eqnarray}}
\newcommand{\eea}{\end{eqnarray}}

\usepackage{soul}

\DeclareMathOperator{\D}{d}
\DeclareMathOperator{\acosh}{acosh}
\DeclareMathOperator{\acos}{acos}
\DeclareMathOperator{\atanh}{atanh}
\DeclareMathOperator{\atan}{atan}

\begin{document}

\title{Hawking radiation on the lattice from Floquet and local Hamiltonian quench dynamics}
	
	\author{Daan Maertens}
	\affiliation{Department of Physics and Astronomy, University of Ghent, Krijgslaan 281, 9000 Gent, Belgium}
	\author{Nick Bultinck}
	\affiliation{Rudolf Peierls Centre for Theoretical Physics, Parks Road, Oxford, OX1 3PU, United Kingdom}
	\affiliation{Department of Physics and Astronomy, University of Ghent, Krijgslaan 281, 9000 Gent, Belgium}
	\author{Karel Van Acoleyen}
	\affiliation{Department of Physics and Astronomy, University of Ghent, Krijgslaan 281, 9000 Gent, Belgium}

	\begin{abstract}
    We construct two free fermion lattice models exhibiting Hawking pair creation. Specifically, we consider the simplest case of a $d=1+1$ massless Dirac fermion, for which the Hawking effect can be understood in terms of a quench of the uniform vacuum state with a non-uniform Hamiltonian that interfaces modes with opposite chirality. For both our models we find that additional modes arising from the lattice discretization play a crucial role, as they provide the bulk reservoir for the Hawking radiation: the Hawking pairs emerge from fermions deep inside the Fermi sea scattering off the effective black hole horizon. Our first model combines local hopping dynamics with a translation over one lattice site, and we find the resulting Floquet dynamics to realize a causal horizon, with fermions scattering from the region outside the horizon. For our second model, which relies on a purely local hopping Hamiltonian, we find the fermions to scatter from the inside. In both cases, for Hawking temperatures up to the inverse lattice spacing we numerically find the resulting Hawking spectrum to be in perfect agreement with the Fermi-Dirac quantum field theory prediction.
    \end{abstract}
	
 	\maketitle
 	
 \tableofcontents

\section{Introduction}
In 1974, Hawking showed in a seminal paper that quantum effects cause black holes to emit thermal radiation \cite{HawkingNature}. This surprising and intriguing result was originally derived \cite{Hawking}, and has since then mostly been discussed, in the framework of quantum field theory (QFT) in classical curved spacetime \cite{Davis,Fulling,Wald}. Following Hawking's result, Unruh showed that a sonic horizon in a fluid will similarly emit thermal radiation \cite{Unruh1981}. This groundbreaking insight gave rise to the field of analog gravity \cite{review1,Gibbons}, and since then different experimental platforms for observing analog Hawking radiation have been put forward, including classical water waves \cite{Weinfurtner}, superfluid Helium\cite{JacobsonVolovik}, ion traps \cite{ciracion} and Bose-Einstein condensates \cite{Cirac}, for which Steinhauer recently reported the actual detection of spontaneously emitted Hawking radiation \cite{Steinhauer2016}. Moreover, on the theoretical front there has been the realization that the analog gravity models bypass the \emph{trans-Planckian problem} \cite{Jacobson1991,Unruh1995,Brout,Jacobson1996}: where for the strictly relativistic QFT case, the Hawking radiation suspiciously relies on an infinite reservoir of arbitrary short distance near-horizon modes, it was shown that this is no longer the case for the Lorentz-violating analog gravity models \cite{Unruh1995,Jacobson1996,Corley,CorleyJacobson, Jacobson_1999,UnruhSchutzhold1,UnruhSchutzhold2,Macher_2009,Parentani}. 

In this work, we approach the phenomenon of Hawking radiation from a purely quantum many-body point of view. In contrast to previous works on Hawking radiation in analog gravity, we consider lattice models with fermionic degrees of freedom (see also Refs.\ \cite{Minar,Celi,Blommaert,LewisVidal,YangLiu,ShiYang}). We also do not try to construct models that exactly emulate the behavior of a quantum field in a black hole spacetime. Instead, we aim to explain the Hawking effect entirely in terms of simple physical concepts associated with quantum many-body systems. In particular, we argue that to understand Hawking radiation, no knowledge of general relativity is required \textemdash it can be understood as a universal quantum dynamics phenomenon in many-body systems with gapless, linearly dispersing excitations. 

\begin{figure}
    \centering
    \includegraphics[scale=0.5]{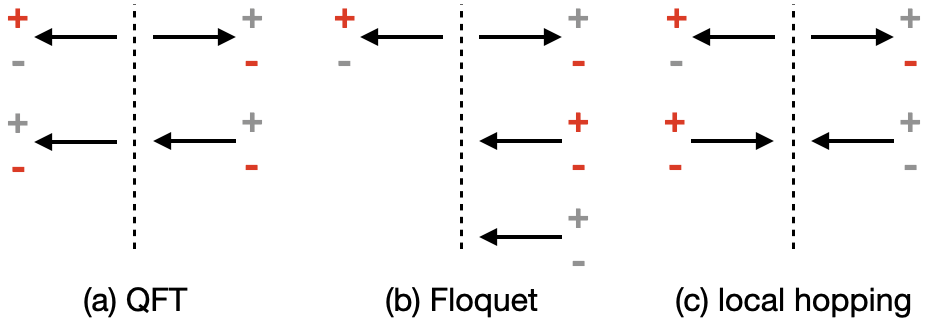}
    \caption{The black hole horizon as a boundary between chirality-changing modes, with the inside of the (effective) black hole on the left. In all cases, the right-mover outside becomes a left-mover on the inside (upper arrows). (a) For a massless Dirac fermion (QFT) the extra left-mover decouples and is unaltered across the horizon; (b) for the Floquet model, we have two additional left-moving modes (``doublers'') on the outside; (c) for the local hopping model, we have a right-moving ``doubler'' on the inside and a left-moving `doubler' on the outside. The + (\textminus) symbols give the occupation of the positive- \mbox{(negative-)} energy states, for the initial ``Minkowski groundstate'' (see text), with grey (red) denoting empty (filled) modes.}
    \label{fig:cartoon}
\end{figure}

To motivate our set-up and terminology let us briefly discuss the Hamiltonian picture of the original Hawking process on curved space-times. The  Hawking radiation results from time evolution with a \emph{time-dependent} Hamiltonian $H(t)$. The time-dependent Hamiltonian is taken to represent the following process. In the infinite past, we assume that there is no black hole, and hence no horizon. There is only an extremely thinly spread mass density. Note that we consider this mass to be classical: the mass density profile acts \textemdash through its effect on the spacetime as dictated by the Einstein equations \textemdash as a background in which the quantum degrees of freedom are living. As such the Hamiltonian evolution at early times, corresponds to a good approximation to that of empty Minkowski space-time. Next, as time progresses, the mass density profile starts to collapse under its own gravitational force. At some finite time $t_c$, the collapse produces a black hole. The Hamiltonian $H(t>t_c)$ thus evolves the quantum degrees of freedom forward in time in the presence of a horizon. A quantum Hamiltonian in the background of a realistic model of gravitational collapse will of course be very complicated. For example, even before the horizon is formed, the concentrated mass density profile will curve space-time which already leads to non-trivial dynamics for the quantum degrees of freedom. Also, right after the horizon is formed it takes some time for transients to die out and the black hole to reach the stationary Schwarzschild (or Kerr) solution. This will also produce non-trivial dynamics for the quantum degrees of freedom. However, one of the most remarkable features of Hawking radiation is that it does not depend on all these details. It also does not depend on the rate at which the gravitational collapse occurs. Hawking radiation solely results from the fact that in the infinite past there is no horizon [and that at $t=-\infty$ we start in the ground state of $H(-\infty)$], and that for some finite $t_c$, the Hamiltonian $H(t>t_c)$ evolves the quantum degrees of freedom in the presence of a horizon. In this paper, we will therefore consider the most crude approximation to the time-dependent Hamiltonian:
\begin{equation}
H(t) = \begin{cases}
      H_0 & \text{if } t<0\\
      H_h & \text{if } t>0
    \end{cases}
    \,,
\end{equation}
where $H_0$ describes time evolution in the \emph{absence} of a horizon, and $H_h$ describes time evolution in the \emph{presence} of a horizon. Our time-dependent Hamiltonian thus describes a ``quench'' process where the Hamiltonian is suddenly switched from $H_0$ to $H_h$ at $t=0$. Note that as we start out in the ground state of $H_0$ ( the ``Minkowski groundstate''), the time evolution at $t<0$ is trivial, and we only have to focus on the time evolution at $t>0$. 

The final remaining question to understand Hawking radiation is what it means for a quantum Hamiltonian to ``have a horizon''. In its most stripped down version, a (static) causal horizon for a free massless Dirac fermion is a boundary between a region with both left- and right-moving modes (right of the horizon) and a region with two left-moving modes (left of the horizon). This is illustrated in Fig.\ \ref{fig:cartoon} (a). The region ``behind'' the horizon corresponds to the chiral region with two left movers. Particles behind the horizon can clearly never cross the horizon. Note that for concreteness, we have made a particular choice of horizon here \textemdash we could equally well have the horizon separate two right-movers from a left- and right-mover.

In both cases, the Hawking effect is entirely due to the mode which changes its chirality upon crossing the horizon. In particular, as we discuss in more detail below, Hawking radiation occurs when the ground state of the Hamiltonian $\hat{H}_0$, which has a left- and right-mover everywhere, is evolved in time with a Hamiltonian which has a chirality-changing mode at the horizon. In Appendix \ref{app:continuum}, we show that for a continuum Dirac fermion this quench process indeed produces thermal Hawking radiation with a temperature $T_H = (2\pi k_B)^{-1} \hbar\partial_x  v(x)\big|_{x=0}$, where $k_B$ is Boltzmann's constant, $x=0$ is the location of the horizon, and $v(x)$ is the spatially-varying velocity of the chirality-changing mode. 

The chiral nature of the region behind the horizon is a problem for lattice models, as the Nielsen-Ninomiya theorem states that every local lattice Hamiltonian has a zero net chirality \cite{NielsenNinomiya}. To circumvent this problem, we adopt two different approaches. In the first, we discretize the time evolution \textemdash turning the dynamics into a Floquet problem. In this case we are able to realize an exact causal horizon on the lattice, albeit at the price of going beyond the framework of local Hamiltonian time evolution. In the second approach, we study continuous time evolution with a local lattice Hamiltonian which does not have a strict causal horizon, but does have a special point where \emph{both} the left- and right-moving modes simultaneously change chirality (such that the net chirality is zero everywhere). For both models, we find the Hawking effect in excellent agreement with the QFT results.

\section{Floquet dynamics}

This section explains the first of our two approaches to realize Hawking radiation in a lattice model, where the chiral dynamics in the region behind the horizon is implemented via a Floquet time evolution operator. Note that in the remainder of the manuscript we will take $\hbar = 1$.

\subsection{The model}

As explained in the introduction, we would like to approximate the time-dependent Hamiltonian $H(t)$ which generates Hawking radiation with a \emph{pair of static} Hamiltonians: $H(t) = \Theta(-t)H_0 + \Theta(t)H_h$, where $\Theta(t)$ is the Heaviside step function and $H_0$ ($H_h$) evolves the quantum degrees of freedom in the absence (presence) of a horizon. However, as was also mentioned in the introduction, the Nielsen-Ninomiya theorem precludes a lattice discretization of $H_h$. To circumvent this issue, in this section we discretize the time evolution in steps of $\Delta t$, and define the time evolution operator which evolves the state at time $n\Delta t$ to the state at time $(n+1)\Delta t$ ($n\in\mathbb{Z})$ as
\begin{equation}
U((n+1)\Delta t; n\Delta t) =
\begin{cases}
e^{-i\Delta tH_0} & \text{ if } n < 0\\
U_F(\Delta t)  & \text{ if } n \geq 0\,,
\end{cases}
\end{equation}
where $U_F(\Delta t)$ is the Floquet time evolution operator defined below. So in this section, we replace the pair of static Hamiltonians with a static Hamiltonian $H_0$ and a Floquet time evolution operator $U_F(\Delta t)$. As explained in the introduction, the initial state at $t = -\infty$ is the ground state of $H_0$. So the time evolution for $t<0$ is trivial, and the sole purpose of $H_0$ is to obtain a corresponding ground state which serves as the initial state for the time evolution.

We start by defining the Hamiltonian $H_0$, which is translationally invariant. It is given by a simple free fermion Hamiltonian describing electrons hopping on a 1D spatial lattice: 
\begin{equation}\label{H0Fl}
H_0 = \sum_{j=1}^N \frac{\gamma}{2}(ic^\dagger_{j+1}c_j -i c_j^\dagger c_{j+1})\,,
\end{equation}
where $c^\dagger_j,c_j$ satisfy the canonical fermion anti commutation relations $\{c^\dagger_j, c_{j'}\} = \delta_{j,j'}$, $\{c^\dagger_j,c^\dagger_{j'}\} = \{c_j,c_{j'}\} = 0$ and we take periodic boundary conditions $c_0=c_N$. This Hamiltonian is readily diagonalized in momentum space, $\hat{H}_0=\sum_k\omega_0(k)\tilde{c}^\dagger_{k}\tilde{c}_k$, with $\tilde{c}_k=\frac{1}{\sqrt{N}}\sum_j e^{-i k j a} c_j$ and $\omega_0(k)=\gamma \sin(ka)$. Note that as we are working on a finite system, the allowed momenta are discrete: $k_m = 2\pi m/(Na)$, where $m \in \{-N/2,-N/2+1,\dots,N/2-1\}$ (we take $N$ to be even). For simplicity of notation we will drop the subscript $m$, and the momenta will always be discrete unless explicitly mentioned otherwise. We fix the chemical potential to zero, such that all single-particle states with negative $\omega_0(k)$ are occupied in the ground state (with zero chemical potential, the lowest-energy state lies in the Fock space sector with $N/2$ particles). The occupied states thus correspond to the single-particle states with negative momenta contained in the interval $[-\pi/a,0]$. The ground state can therefore be written as
\begin{equation}\label{psi0}
\ket{\psi_0}=\prod_{-\pi/a<k<0} \tilde{c}^\dagger_{k}\ket{0}\,,
\end{equation} 
with $c_j\ket{0}=0$. $|\psi_0\rangle$ is the initial state for the time evolution.

Note our somewhat unconventional choice of the fermion hopping terms in Eq. \eqref{H0Fl}, with the $\pm i$ factors. With this choice we can identify the right-mover of the Minkowski QFT, with dispersion $\omega_0(k)=c k$, with the \emph{gapless/massless} modes of the lattice model near $k\approx 0$, where we have $\omega_0(k)=\gamma \sin{ka}\approx v k$. Here gappless/massless refers to the fact that their excitation energy can be made arbitrary small. This means that we can make wavepackets on top of the ground state $|\psi_0\rangle$ constructed from plane wave states with arbitrarily (in the thermodynamic limit) small energies and momenta which will be right-moving under forward time evolution, i.e., they have a positive group velocity $+v$. Similarly, the other gapless excitation located around $k\approx -\pi/a$, with $\omega_0(-\pi/a+\tilde{k})\approx - v\tilde k$ for small $\tilde k$, can be identified with a QFT left-mover. As mentioned in the introduction, the Hawking effect is due to a massless excitation whose group velocity changes sign behind the horizon, $\omega(k)=+c k \rightarrow \omega(k)=-c' k$, once the horizon is formed. As we will show, for both our lattice models, it are precisely the modes near $k\approx 0$ that play this role.   

Next we define the Floquet time evolution operator $U_F(\Delta t)$ which time-evolves the fermions in the presence of a horizon. The Floquet unitary can be written as a product of two operators:

\begin{equation}\label{FlU}
    U_F(\Delta t) = \hat{T}_L e^{-i\Delta t H}\,,
\end{equation}
where 
\begin{equation}\label{HFl}
H = \sum_{j=1}^N \frac{\gamma_j+\gamma_{j+1}}{4}(ic^\dagger_{j+1}c_j -i c^\dagger_j c_{j+1})
\end{equation}
is a generalization of $H_0$ which is not translationally invariant, and $T_L c^\dagger_j T_L^{-1} = c^\dagger_{j-1}$ (with $c^\dagger_0 = c^\dagger_N$) implements a translation to the left. Before explaining in detail how the $\gamma_j$ in Eq. \eqref{HFl} are chosen, let us first elaborate on why $U_F(\Delta t)$ represents a Floquet time evolution operator. $U_F(\Delta t)$ first time-evolves the system over one period $\Delta t$ with $H$, after which it is instantaneously ``kicked'' by acting with $T_L$. Experimentally, $T_L$ would be implemented by physically translating the system over one lattice site, similarly to the rotating ion trap proposal of Ref. \cite{ciracion}. This sequence is then repeated indefinitely, such that the time evolution is periodic in time. In this work we will limit ourselves to a ``stroboscopic'' description of the system, where only states at discrete times $n\Delta t$ are considered. In this case, it is sufficient to exclusively work with the time-independent evolution operator $U_F(\Delta t)$. The latter is in general defined as the Floquet time evolution operator in periodically-driven systems.

\begin{figure}
    \centering
    \includegraphics[scale=0.4]{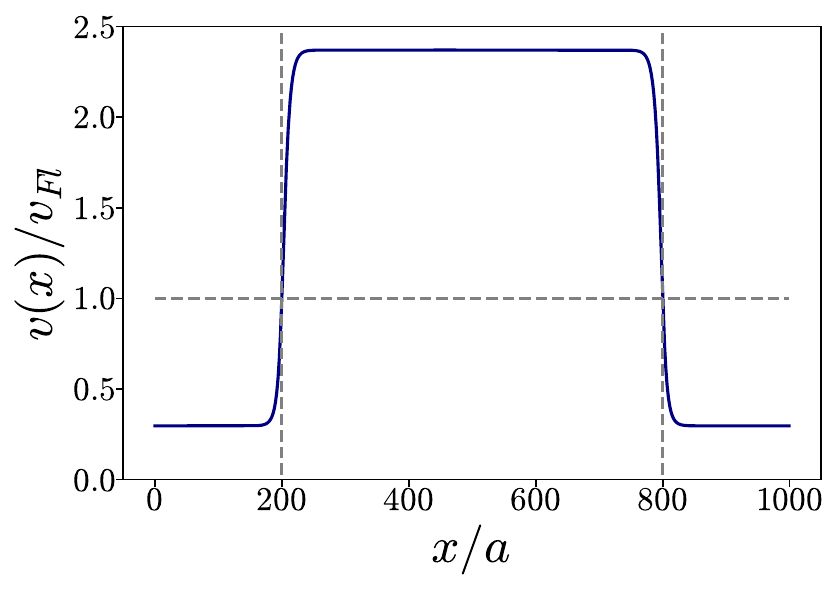}
    \caption{Example of $v(x)$ with $\kappa  \Delta t= 0.1 $. (Eq.\ \eqref{tx} in Appendix \ref{app:Floquet} with $N=L/a=1000$, $W/a = 600$, $b = 3$, and $\tilde{\kappa} a=0.1 $.) The locations of the black hole horizon $x_b = 200 a$ and white hole horizon $x_w = 800 a$ are indicated with dashed lines.}
    \label{fig:vprofile}
\end{figure}

The spatially-varying hopping strengths in Eq. \eqref{HFl} are defined as $\gamma_j = \gamma(ja)$, where $\gamma(x)$ is a continuous function of position. Associated with $\gamma(x)$, we can define a corresponding velocity function $v(x) = a\gamma(x)$. Defining the intrinsic Floquet velocity as $v_{Fl} = a/\Delta t$, the velocity function satisfies $v(x) > v_{Fl}$ for $x_{b} < x < x_{w}$, and $v(x) < v_{Fl}$ for $x < x_{b}$ and $x > x_{w}$, where $x_{b}$ and $x_{w}$ respectively denote the location of a black hole horizon ($\partial_x v(x)|_{x=x_b} > 0$) and a white hole horizon ($\partial_x v(x)|_{x=x_w} < 0$). Note that we are forced to simultaneously introduce a black hole and a white hole horizon because of the periodic boundary conditions. In Fig.\ \ref{fig:vprofile}, we show an example of $v(x)$ used in our numerics. From the properties of $v(x)$, it is intuitively clear why the Floquet unitary in Eq.\ \eqref{FlU} has a black hole horizon at $x = x_b$. During the first part of the Floquet time evolution, a particle at position $j$ can travel a distance $v(ja)\Delta t$ to the right. After this initial time evolution, the Floquet unitary implements a translation to the left over one lattice constant $a$. If $v(ja)\Delta t < a$, or equivalently $v(ja) < v_{Fl}$, then every particle necessarily has a net displacement to the left during one Floquet time-step. This is true for particles behind the black hole horizon, corresponding to the region with $ja<x_b$ and $ja>x_w$. Notice that $v_{Fl}$ and $v(x)$ are respectively the analog of the fluid velocity and sound mode velocity for a sonic black hole \cite{Balbinot_2008}. The white hole horizon at $x_w$ corresponds to a point from which particles can only escape, but not enter, the region behind the black hole horizon. In this work, we focus on the dynamics resulting from the black hole horizon, and from now on we will refer to the black hole horizon simply as the horizon, and the region $x_b<ja<x_w$ as the region outside the horizon. 

As a final comment, let us note that $\hat{T}_L$ cannot be written as the exponential of a local Hamiltonian \cite{Gross,Cirac}. This puts our model outside the well-studied class of 1D dynamics models consisting of finite-depth unitary circuits. However, our model can be realized on the edge of a 2D finite-depth quantum circuit \cite{Demler,Po,Fidkowski}, or by having a moving horizon and working in the co-moving frame, as in Refs.\ \cite{ciracion,Steinhauer2016}. This latter approach was also considered in the bosonic lattice model of Corley and Jacobson \cite{CorleyJacobson}. See Appendix \ref{app:CJ} for a detailed discussion of the connection between the Corley-Jacobson model and the Floquet model used in this work.

\subsection{Dynamics of wavepackets far away from the horizon}
Before delving into the numerical study of the time evolution with $U_F(\Delta t)$, it is instructive to first examine the wavepacket dynamics in both the inside and outside regions of the black hole, far away from the horizons. As the Floquet time evolution operator is homogeneous far away from the horizons, it is sufficient to consider two translationally invariant Floquet unitaries in order to understand the dynamics of wavepackets localized in those regions. One translationally invariant Floquet unitary describes wavepacket dynamics in the the inside region, and the other one describes dynamics in the outside region. In conventional Hamiltonian systems describing free particles, wavepacket dynamics is most easily understood by obtaining the dispersion relation through diagonalization of the single-particle Hamiltonian. Our Floquet time evolution, however, is not generated by a simple time-independent Hamiltonian. It is therefore easier to directly determine the single-particle states which are stationary under the stroboscopic time evolution. These single-particle states, with corresponding creation operators $\sum_j A_j(\omega) c^\dagger_j$, are defined to satisfy following relation:
\begin{equation}\label{eq:Fldis}
U_F(\Delta t) (\sum_j A_j(\omega) c^\dagger_j) U_F(\Delta t)^\dagger= e^{-i\omega\Delta t}(\sum_j A_j(\omega) c^\dagger_j)\, ,
\end{equation}
To study the dynamics of wavepackets in the translationally invariant regions far away from the horizon, we can take $U_F(\Delta t)$ to be translationally invariant, in which case one can obtain a ``Floquet dispersion relation'', i.e., $\omega$ as a function of momentum $k$. The Floquet dispersion relation obtained by using $U_F(\Delta t)$ as defined in Eq. \eqref{FlU}, with $H$ replaced by $H_0$ defined in Eq. \eqref{H0Fl}, is
\begin{equation}\label{omegak}
\omega(k) = v\sin(ka)/a - kv_{Fl} \text{ mod } 2\pi/\Delta t\,,
\end{equation}
where we have emphasized that $\omega$ defined in Eq. \eqref{eq:Fldis} is only determined modulo $2\pi/\Delta t$. The spectral shift by $kv_{Fl}$ in Eq. \eqref{omegak} is exactly the same as the one which occurs by going from the background fluid frame to the laboratory frame in continuum fluid models that exhibit the Hawking effect. In fact, that such a spectral shift allows for the realization of sonic horizons was part of Unruh's original insight that lead to the field of analog Hawking radiation \cite{Unruh1981}.

\begin{figure}
    \centering
    \includegraphics[scale=0.25]{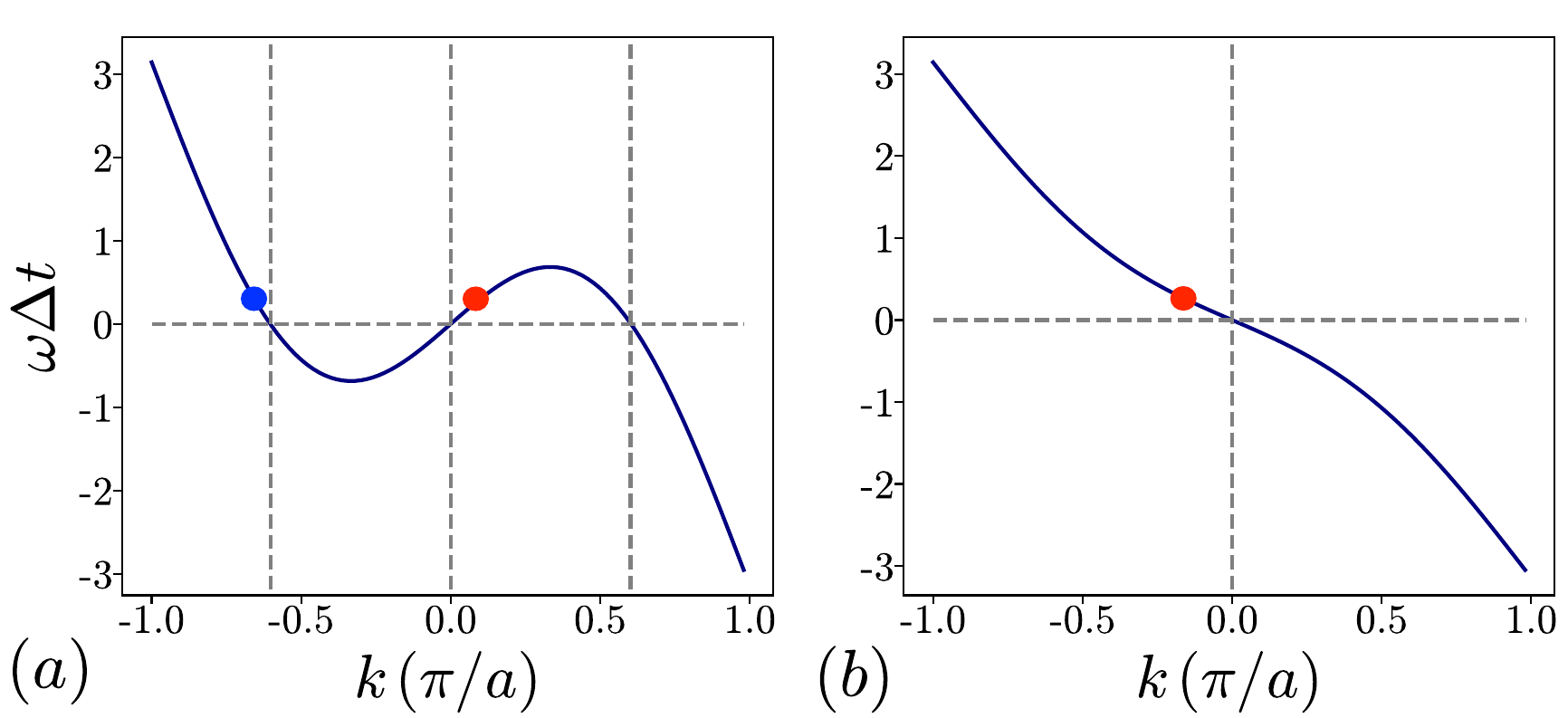}
    \caption{[(a) and (b)] Floquet frequencies for spatially uniform $\gamma_j = \gamma$. In (a), $v = at = 2v_{Fl}$. This corresponds to the region outside the horizon. The vertical dashed lines show $k=0$ and $k = \pm k^*$. In (b), $v = v_{Fl}/2$. This corresponds to the region inside the horizon. The dots situate the modes involved in the forward scattering process (see main text).}
    \label{fig:Fldisps}
\end{figure}

The Floquet dispersion relation $\omega(k)$ is shown both for the case where $v > v_{Fl}$ [Fig.\ \ref{fig:Fldisps}(a)], and the case where $v<v_{Fl}$ [Fig.\ \ref{fig:Fldisps}(b)]. The former corresponds to the region outside the region, and the latter to the region behind the horizon. Note that because Floquet frequencies are only defined mod $2\pi/\Delta t$, $\omega(k)$ is a continuous function of $k$ over the entire Brillouin zone. We see that in the outside region, wavepackets with small momenta are right-moving (because of the positive group velocity near $k=0$ obtained from $\omega(k)$), whereas the same wavepackets are left-moving on the inside. These wavepackets thus represent the low-energy excitations with changing chirality, or equivalently, sign-changing group velocity, upon crossing the horizon which are essential for the realization of a horizon, as explained in the introduction. Also note that outside the horizon [Fig.\ \ref{fig:Fldisps}(a)], $\omega(k)=0$ not only at $k=0$, but also at two other non-zero momenta $k = \pm k^*$, meaning that right-moving wavepackets constructed from small momenta will be degenerate with left-moving wavepackets constructed from momenta near $\pm k^*$. We call these degenerate left-moving wavepackets the two `doubler' modes, which are also shown in Fig.\ \ref{fig:cartoon}(b). The doubler modes will play an important role in our interpretation of the numerical results presented below. From now on, we will also refer to $\omega$ with $|\omega|\ll 2\pi /\Delta t$ as the `energy' of wave packets.

As a final comment, let us note that obtaining all stationary single-particle states which satisfy Eq. \eqref{eq:Fldis} with the original definition of $U_F(\Delta t)$ (i.e., using the non translation invariant $H$) provides a complete solution of the problem, and in particular for the time evolution that we are interested in here. In fact, this is exactly how we obtained the numerical results presented in the next section.

\subsection{Numerical results}\label{sec:floquet}

Hawking radiation emerges upon quenching the ground state of $\hat{H}_0$ with $U(\Delta t)$. To measure the Hawking radiation we define the wave packet creation operators $W^\dagger_{x_0,\omega}$, where $x_0$ is the location of the wave packet in real-space, and $\omega$ its energy [see Eq. (\ref{eq:wavepacket})]. Importantly, the wave packets $W^\dagger_{x_0,\omega}$ only contain momenta near $k = 0$ such that they are right-moving on the outside of the horizon. The time-dependent occupation number of the wave packets is given by

\begin{equation}\label{wpocc}
    N_{x_0,\omega}(t) = \langle \psi_0(t)|W^\dagger_{x_0,\omega}W_{x_0,\omega}|\psi_0(t)\rangle\, ,
\end{equation}
with $t = n \Delta t$, 
\begin{equation}
|\psi_0(n\Delta t)\rangle = U^n(\Delta t) |\psi_0\rangle\,,
\end{equation}
and $|\psi_0\rangle$ is the ground state of $\hat{H}_0$ defined in Eq \eqref{psi0}.

\begin{figure}
    \centering
    \includegraphics[scale=0.39]{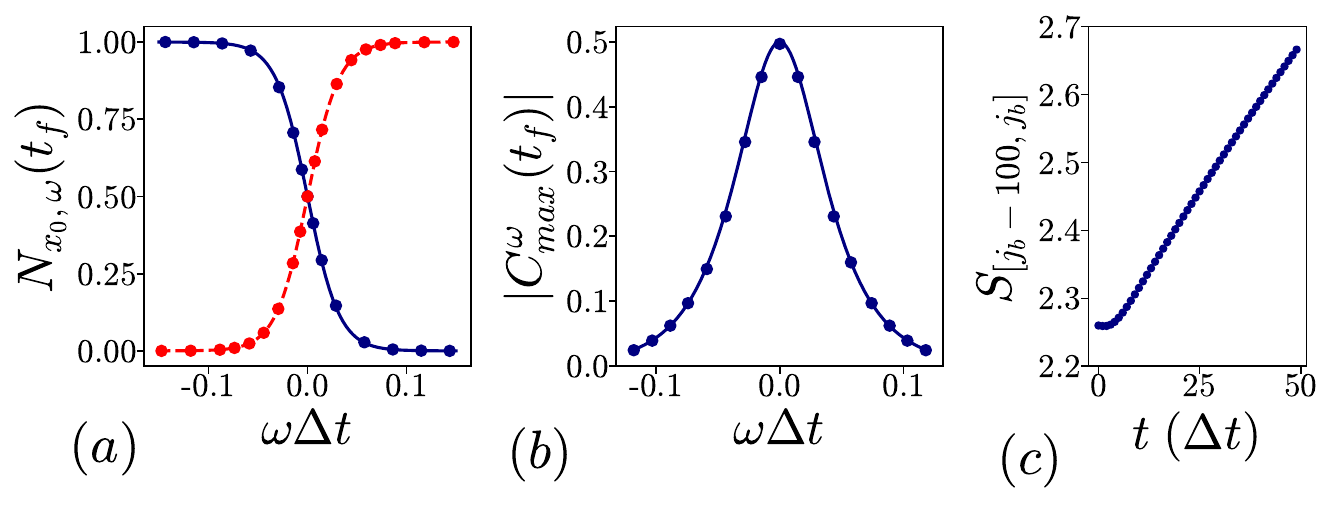}
    \caption{Numerical results obtained using $N=3000$, $j_b = 500$, $j_w = 2500$ and $\kappa \Delta t= 0.1$ (for more details, see Appendix \ref{app:Floquet}). (a) wave packet occupation numbers at time $t_f$ and position $x_0$. Blue points correspond to $x_0/a = j_b + 700$ and $t_f = 1000 \Delta t$. The blue full line is $f(\omega)$. Red points correspond to $x_0/a = j_b-450$ and $t_f = 1200 \Delta t$. The red dashed line is $f(-\omega)$. (b) Maximal wave packet correlation $|C_{max}^\omega(t_f)| = \text{max}_j| C^{\omega}_{j_b - j_{in},j_b+j}(t_f)|$ across the horizon with $j_{in} = 450$ and $t_f = 1000\Delta t$. The full blue line is $\sqrt{f(\omega)f(-\omega)}$. (c) Entanglement entropy $S_{[j_b-100,j_b]}$ of the spatial interval $[j_b-100,j_b]$ as a function of time.}
    \label{fig:FLresults}
\end{figure}

The initial wave packet occupation number $N_{x_0,\omega}(0)$ for $x_0 \gg x_b$ and $x_0\ll x_w$ is a step function $\Theta(-\omega)$, smeared over the width of the wave packet in energy space. In full agreement with the QFT case, our lattice simulations show that the quench produces outgoing particles near the horizon, with velocity $v_{out}=v(x_0)-v_{Fl}$ (considering  $v(x)$ to be approximately constant away from the horizon). In particular, we find that after a time $t^* \sim |x_0-x_b|/v_{out}$ the wave-packet occupation number changes to a distribution that has an excellent fit to a Fermi-Dirac distribution:

\begin{equation}\label{eq:FD}
    N_{x_0,\omega}(t \gtrsim t^*) = \frac{1}{e^{\omega/k_B T_H}+1} =: f(\omega)\, ,
\end{equation}
with Hawking temperature 
\begin{equation}
T_H = \frac{\kappa}{2\pi k_B}\, , 
\end{equation}
where $\kappa = \partial_x v(x)|_{x=x_b}$ is the `surface gravity' of the horizon. In Fig.\ \ref{fig:FLresults}(a), we show the numerically obtained Fermi-Dirac distribution for the wave packets using $\kappa \Delta t = 0.1 $ (see Appendix \ref{app:Floquet} for the details of our numerical simulations). At larger times $t\gtrsim t^*+|x_w-x_b|/v_{out}$, the white hole horizon starts to affect the occupation number, and deviations from the Fermi-Dirac distribution set in.

We also calculated the occupation number of wave packets on the inside of the horizon, which we now find to go from an initial smeared out step-function $\Theta(+\omega)$ to a Fermi-Dirac distribution with negative temperature $-T_H$, at times $t \gtrsim t^* \sim |x_0-x_b|/v_{in}$, with $v_{in}=v_{Fl}-v(x_0)$. In Fig.\ \ref{fig:FLresults}(b), we plot 
\begin{equation}
|C^\omega_{max}(t_f)| = \text{max}_{j}| C^{\omega}_{j_b - j_{in},j_b+j}(t_f)|\,, 
\end{equation}
where $j_b = \lfloor x_b/a \rfloor$ and 
\begin{equation}
C^{\omega}_{ij}(t) = \langle \psi_0(t)|W^\dagger_{i,\omega} W_{j,\omega}|\psi_0(t)\rangle
\end{equation}
measures correlations between wave packets with equal energies at different locations. We find that the maximal wave packet correlation across the horizon occurs for $j= j_{in} (v_{out}/v_{in})$, and can almost perfectly be fitted with $|C^\omega_{max}(t_f)| = \sqrt{f(\omega)f(-\omega)}$, again in nice agreement with the QFT result (Eq. \ref{cross}). In Fig.\ \ref{fig:FLresults}(c), we plot the entanglement entropy of the spatial interval $[j_b - 100,j_b]$. It shows a linear increase of entanglement across the horizon.

\begin{figure}
    \centering
    \includegraphics[scale=0.34]{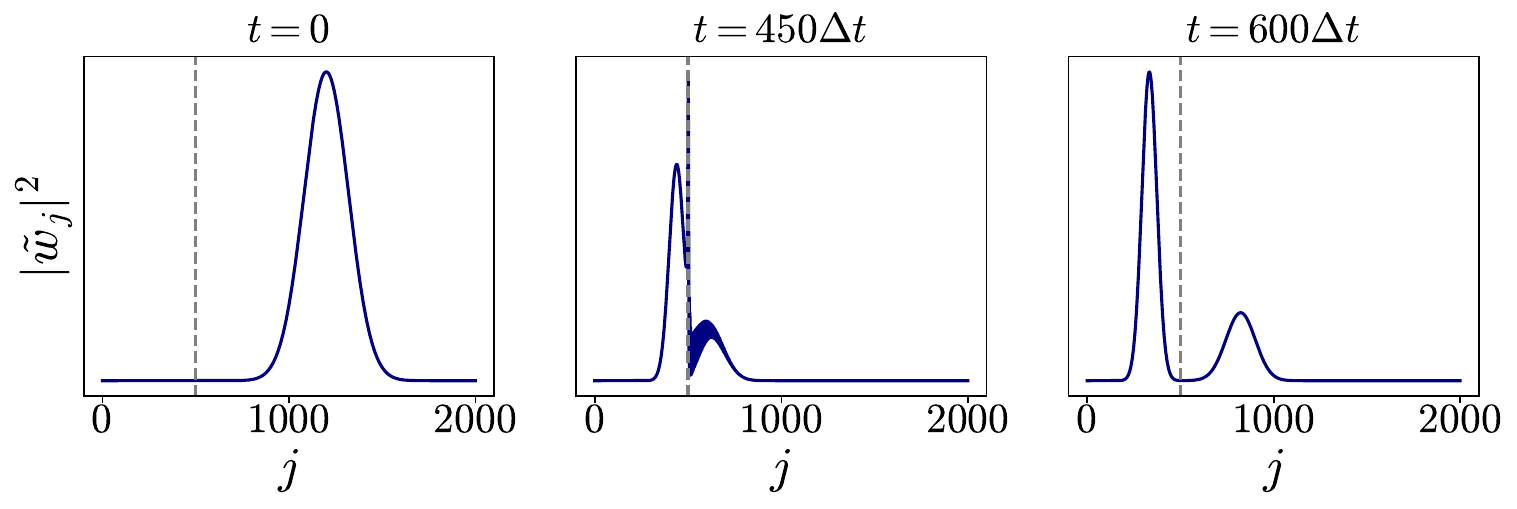}
    \caption{Time evolution of the wave packet $\tilde{W}^\dagger_{x_0,\omega} = \sum_j \tilde{w}_j c^\dagger_j$ made from momenta near $-k^*$, with $x_0/a = j_b + 700$ and $\omega\Delta t = 0.014$. The dashed line is the black hole horizon, and $\kappa \Delta t  = 0.1$ was used.}
    \label{fig:WPO}
\end{figure}

Our numerical results can be understood via the following physical picture. On the outside of the black hole horizon, the wave packets $\tilde{W}^\dagger_{x_0,\omega}$ constructed from momenta near $-k^*$ are occupied in the Minkowski ground state and are moving to the left, i.e. towards the horizon. In Fig.\ \ref{fig:WPO}, we show the time evolution of such a wave packet. We see that part of it is transmitted through the horizon, and part of it is reflected back. The reflected part corresponds to the wave packets $W^\dagger_{x_0,\omega}$ whose occupation number is being measured in Eq.\ \eqref{wpocc}. For clarity, we have marked the parts of the Floquet dispersion corresponding to the incoming (reflected and transmitted) wave packets in this scattering process with a blue (orange) dot in Figs.\ \ref{fig:Fldisps}(a) and \ref{fig:Fldisps}(b). From our numerics, we find that the incoming left-moving wave packets $\tilde{W}^\dagger_{x_0,\omega}$ are transmitted through the horizon with probability $f(-\omega)$, and are reflected back with probability $f(\omega)$. After scattering, the wave packet states thus add a contribution $S = -f(\omega)\ln f(\omega) -f(-\omega)\ln f(-\omega)$ to the entanglement entropy across the horizon. This gives rise to a linear growth of the entanglement entropy, in accordance with the general picture for entanglement growth during quench dynamics put forward by Calabrese and Cardy \cite{CardyCalabrese}. 

To summarize, we find that Hawking radiation is the result of electrons deep inside the Fermi sea of the Minkowski ground state scattering off the horizon. As such, this realizes a fermionic lattice version of the subluminal scenario of \cite{Corley,UnruhSchutzhold1}. Notice also that the scattered single-particle states (right panel in Fig.\ \ref{fig:WPO}) can be interpreted as particle/hole  Hawking-pairs, if we consider them as excitations on top of the initial Minkowski-state $\ket{\psi_0}$. For example, for $\omega >0$, we can write  
\begin{align}
    \ket{\psi_0(t)}&\sim\left(\sqrt{f(-\omega)} W^\dagger_{x_{in} ,\omega}+\sqrt{f(\omega)} W^\dagger_{x_{out},\omega}\right)\ket{0} \nonumber\\
    & \propto\left(1+e^{-\pi\omega/\kappa} W^\dagger_{x_{out},\omega}W_{x_{in},\omega}\right) W^\dagger_{x_{in},\omega}\ket{0}\,,
\end{align}
where in the last line we have dropped a proportionality constant $\sqrt{f(-\omega)}$, and we have used that $\sqrt{f(\omega)/f(-\omega)} = e^{-\pi\omega/\kappa}$.

\section{Local Hamiltonian dynamics} 

After having analyzed the Floquet model with a strict causal horizon in the previous section, we now study a different type of dynamics which is generated by a local Hamiltonian, and hence has no strict horizon, but nevertheless admits Hawking radiation.  

\subsection{The model}
\begin{figure}
    \centering
    \includegraphics[width=\linewidth]{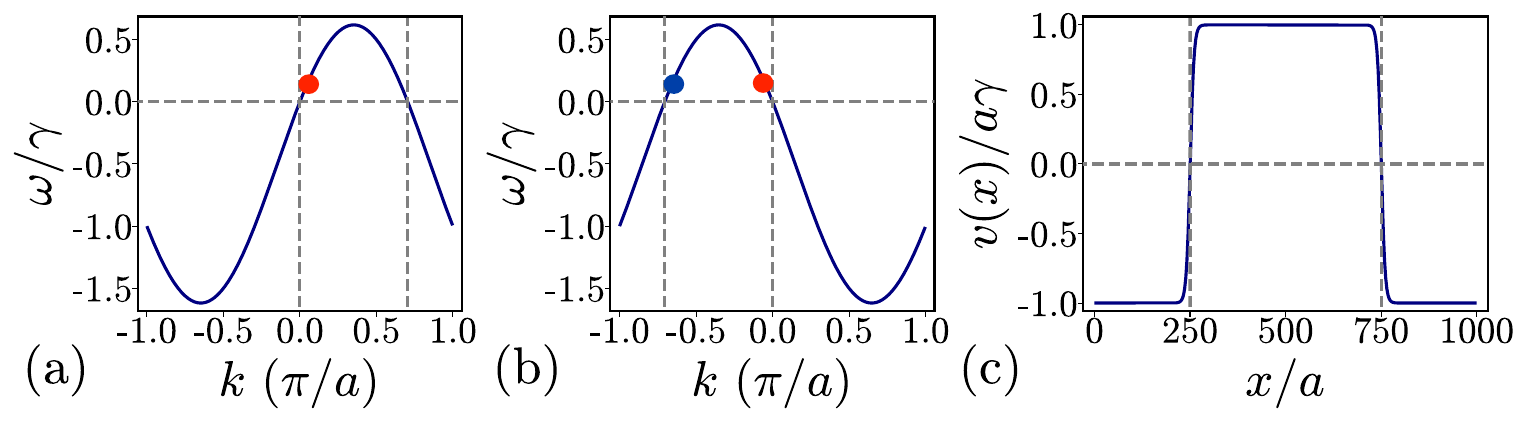}
    \caption{[(a) and (b)] Dispersion relation of the local quench Hamiltonian, far away from the horizon for $\mu=0.5 t$. In (a), $t_j=t$, corresponding to the outside region. In (b), $t_j=-t$, corresponding to the inside region. The vertical lines show the gapless $k=0$ mode and doubler modes: $k=k^*_{out}$ and $k=k^*_{in}$. The blue (orange) dots mark the incoming (reflected/transmitted) modes involved in the forward scattering process of Fig.\ \ref{fig:Snapshot_local}. (c) The spatially-varying velocity profile $v(x) = at(x)$ [Eq.\ \eqref{eq:v_localH} in Appendix \ref{app:local} with $L/a=1000$ and $\hat{\kappa}=0.1$]. The locations of the two boundaries between the inside and outside regions at $x_b=250a$ and $x_w=750a$ are indicated with dashed lines.
    }
    \label{fig:Freqs_local}
\end{figure}

We consider the following local horizon Hamiltonian:
\begin{equation}\label{local}
\hat{H}=\frac{1}{2}\sum_{j=1}^N i\gamma_j c^\dagger_{j+1}c_j  +\mu c^\dagger_{j+1}c_j - \mu c^\dagger_{j}c_j + h.c.\,,
\end{equation}
where the site-dependent hopping term $\gamma_j = \gamma(ja)$ interpolates smoothly (on the lattice scale) between a constant value $-\gamma<0$ in the region $j\lesssim j_b$ and $j\gtrsim j_w$, which represents the analog of the region inside the black hole, to a constant value $\gamma>0$ in the region $j_b\lesssim j \lesssim j_w$ representing the analog of flat space outside the black hole. The specific $\gamma(x)$ profile that we have used in our numerics is shown in Fig.\ \eqref{fig:Freqs_local} (c). 

When $\gamma_j=\pm \gamma$ is constant the Hamiltonian is readily diagonalized in the Fourier basis, resulting in the dispersion relation: \be \omega(k)= \pm \gamma\sin(k a) +\mu \cos(k a) - \mu\,. \ee In Figs.\ \ref{fig:Freqs_local}(a) and \ref{fig:Freqs_local}(b), we show the dispersion relation for the quench Hamiltonian using $\gamma_j = \gamma$, representing the outside region, and $\gamma_j = -\gamma$, representing the inside region. 

Note that, as was already mentioned, there is again a gapless mode at $k = 0$ which is right-moving on the outside and left-moving on the inside. As we are working with a strictly local model, there is now also an additional gapless right-moving mode in the inside region at $k=k^*_{in}<0$ (which is occupied in the Minkowski ground state), and an additional left-moving gapless mode at $k=k^*_{out}>0$ in the outside region (which is unoccupied in the Minkowski ground state). These additional modes correspond to the ``doubler'' modes in Fig.\ \ref{fig:cartoon}(c).

The terms in the Hamiltonian proportional to $\mu$ generate a dispersion $-\mu a^2k^2/2$ at small $k$, so they do not contribute to the velocity of the zero momentum mode. Nevertheless, these terms are crucial because without them, the inside ($\gamma_j<0$) and outside ($\gamma_j>0$) regions are only weakly coupled. We find that for the Hawking effect to occur, the value of $\mu$ should be larger than the surface gravity $\kappa = a\partial_x \gamma(x)|_{x=x_b}>0$. The precise value of $\mu$ does not matter, and we will take $\mu = 0.5 \gamma$. We note that our Hamiltonian differs from previously studied lattice models, obtained by discretizing a Dirac fermion in a black hole spacetime \cite{YangLiu,ShiYang}, exactly by the terms proportional to $\mu$. Also note that the third term in \eqref{local} does not contribute to the dynamics (on particle-number eigenstates). 

For this strictly local model, the doubler modes prevent the boundary between the in- and outside region from acting like a true horizon. Instead, the inhomogeneous region can be viewed as an imperfect filter, letting some modes through while trapping others. Indeed, if we approximate the hopping coefficient inside this region by its minimal value $\gamma_{min}=0$ then the dispersion relation there only supports modes with energy $\omega \in [-2\mu,0]$. This prevents modes with $\omega \gtrsim 0$ (and $\omega \lesssim -2\mu$) from propagating through this region. As a result these modes are effectively trapped inside the homogeneous regions. By quenching the Minkowski ground state with the inhomogeneous Hamiltonian we essentially create a meta-stable state where the trapped modes can tunnel through the region where $\gamma_j$ becomes zero. It is precisely this ``leakage" that we observe as Hawking radiation. In this regard our setup is very similar to that of Hawking emission in superfluids \cite{JacobsonVolovik} and from photon emission out of an optical cavity \cite{Rosenberg_2020}.

\subsection{Numerical results}\label{sec:local}

\begin{figure}
    \centering
    \includegraphics[width=\linewidth]{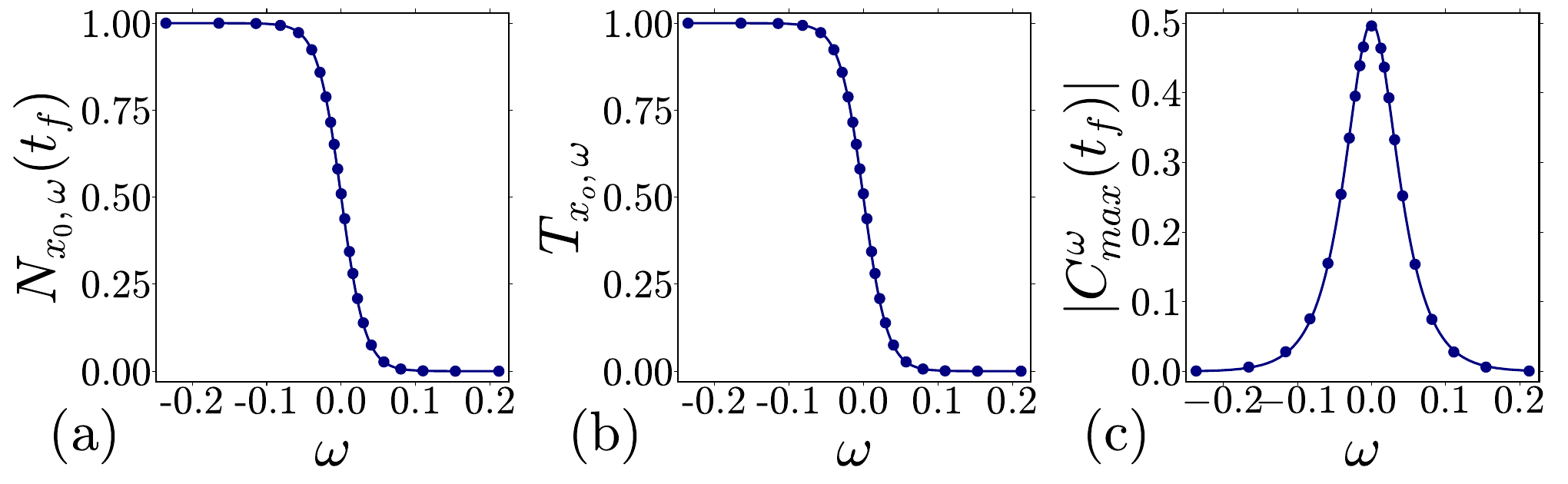}
    \caption{Numerical results using $N=4000$, $j_b = 1700$, $j_w = 3900$, and $\kappa= 0.1 \gamma$. (a) Wave packet occupation numbers at time $t_f$ and position $j_0=j_b+900$ as calculated from \eqref{wpocc}. (b) Spectrum obtained from the transmission coefficient for the time evolution of right-moving wave packets starting at the inside of the horizon $j_0=j_b-900$ (see Fig.\ \ref{fig:Snapshot_local}). The results (a) and (b) are nearly identical and coincide perfectly with $f(\omega)$ (shown by the full line). (c) Wave-packet correlation $| C^{\omega}_{max}(t_f)|=| C^{\omega}_{j_b - 900,j_b+900}(t_f)|$ across the horizon with $t_f = 1850 t^{-1}$, the full line shows $\sqrt{f(\omega)f(-\omega)}$.}
    \label{fig:Occs_local}
\end{figure}

\begin{figure}
    \centering
    \includegraphics[width=\linewidth]{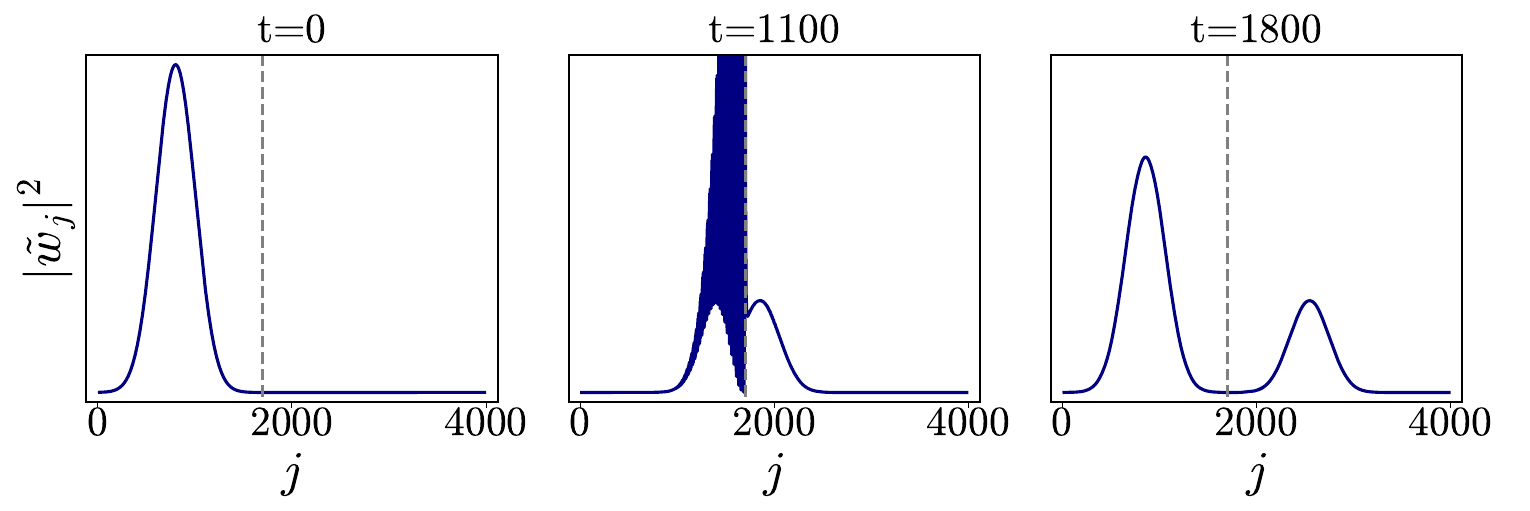}
    \caption{Time evolution of a right-moving wave packet $\tilde{W}^\dagger_{x_0,\omega} = \sum_j \tilde{w}_j c^\dagger_j$ starting in the inside region at $x_0/a = j_b-900$ with $\omega=0.0157$, $\mu=0.5$ and $\kappa =0.1$, expressed in units $t=\hbar =1$ . The dashed line shows the boundary between inside and outside regions. 
    }
    \label{fig:Snapshot_local}
\end{figure}

To measure the Hawking radiation in our local model we first follow the same procedure as in the Floquet case: we quench the groundstate of $\hat{H}_0$ with $U(t)=e^{-it\hat{H}}$, and detect the Hawking particles by measuring the wave packet occupation number as in Eq.\ \eqref{wpocc}. The occupation number spectrum obtained this way shows again 
excellent agreement with the thermal Hawking distribution, as can be seen from Fig.\ \ref{fig:Occs_local}(a-c). However, different from the Floquet model, the Hawking pairs now arise from the filled Dirac-sea of right-moving modes near $k=k^*_{in}$ on the inside. These causality-violating modes now realize a fermionic version of the superluminal scenario of \cite{Corley, UnruhSchutzhold1}. In Fig.\ \ref{fig:Snapshot_local}, we show the time evolution of such a right-moving wave packet that starts out in the inside region at $t=0$.  

For a wave packet of energy $\omega$, we find that the transmission and reflection coefficients are respectively fitted very nicely by the Fermi-Dirac expression $f(\omega)$ and $f(-\omega)$, which again explains both the thermal nature and the correlations of the Hawking pairs.  Similar to the Floquet case, we limit ourselves here to times $t<|x_w-x_b-L|/v_{in}$, before any effects of the white hole horizon in superluminal scenarios (see Refs.\ \cite{BHlaser}) can be measured. Notice that for large values of $\omega\gtrsim \gamma^2/\mu$, we do find deviations from the Fermi-Dirac distribution, see Sec. \ref{subsecD}.

\subsection{Analysis in terms of stationary scattering states}\label{sec:scat}

Instead of investigating the time evolution of wave packets, the same results can also be obtained by constructing stationary scattering states \cite{Corley,CorleyJacobson, Jacobson_1999,UnruhSchutzhold1,UnruhSchutzhold2,Macher_2009,Parentani}. The Hawking temperature is then contained entirely in these stationary scattering states. The only information needed about the initial Minkowski ground state is which momentum modes are occupied. 

Consider the quench Hamiltonian \eqref{local} in the thermodynamic limit ,i.e., $j\in]-\infty,+\infty[$. First recall that for constant $\gamma_j=\gamma$ the Hamiltonian is readily diagonalized with dispersion relation (now taking the momentum $k$ in lattice units, $-\pi<k\leq\pi$) 
\begin{equation}
    \omega_\gamma(k)=\gamma\sin(k) +\mu \cos(k) - \mu\,,
\end{equation}
as shown in Fig.\ \ref{fig:Freqs_local}(a). Let us now take $\gamma_j$ to interpolate between a constant value $\gamma_{-\infty}=-\gamma$ and  $\gamma_{\infty}=\gamma$. More precisely we take $\gamma_{j<j_L}=-\gamma$, $\gamma_{j>j_R}=\gamma$, and the interpolating region to be within $[j_L,j_R]$. For our numerics we used
\begin{equation}
    \gamma_j/\gamma = 1- \frac{2}{1+e^{2\hat{\kappa}j}} = \tanh \hat{\kappa}j\, .
\end{equation}
The operators
\begin{equation}
    \tilde{c}_\omega^\dagger = \sum_j f_\omega(j) c_j^\dagger
    \label{eq:discrete_ansatz}
\end{equation}
correspond to single-particle eigenstate creation operators with energy $\omega$, provided that the coefficients $f_\omega(j)$ satisfy the following Schrödinger-like equation:
\begin{align}
    \frac{-i\gamma_j+\mu}{2} f_\omega(j+1) +\frac{i\gamma_{j-1}+\mu}{2} f_\omega(j-1) = (\omega+\mu) f_\omega(j)\,.
    \label{eq:Schrodingerlike}
\end{align}
This discrete Schrödinger equation \eqref{eq:Schrodingerlike} allows us to define a probability current $J_j$
\begin{align}
    J_j = \frac{\gamma_j}{2} &[f_\omega^*(j)f_\omega(j+1)+f_\omega^*(j+1)f_\omega(j)] \nonumber\\ &+ \frac{i\mu}{2}[f_\omega^*(j)f_\omega(j+1)-f_\omega^*(j+1)f_\omega(j)]\, ,
    \label{eq:cons_current}
\end{align}
which is conserved, i.e. does not depend on $j$: $J_j = J$. 

Equation \eqref{eq:Schrodingerlike} can be be solved inductively: from $f_\omega(j_0+1)$ and $f_\omega(j_0)$ one can solve for $f_\omega(j_0-1)$ and subsequently obtain $f_\omega(j)$ for all $j\in]-\infty,+\infty[$. As such, this equation will have a solution for every $\omega$,
but only for certain $\omega$ the solution is bounded for $j\to \pm\infty$. For a general hopping profile $\gamma(x)$ interpolating between $-\gamma$ and $\gamma$, we anticipate plane wave solutions away from the interpolation region $j\in [j_L,j_R]$:
\begin{equation}\label{eq:planewaves}
    f_\omega(j) = \begin{cases} A_L e^{ik_1j} + A_R e^{ik_2j} ,\quad j>j_R \\ B_L e^{iq_1j} + B_R e^{iq_2j} ,\quad j<j_L \end{cases}
\end{equation}
with $\omega=\omega_\gamma(k_1)=\omega_\gamma(k_2)=\omega_{-\gamma}(q_1)=\omega_{-\gamma}(q_2)$. Here $k_2$ is the momentum of the gapless right-moving mode around $k=0$ and $k_1$ is the momentum of the left-moving mode around $k_{\text{out}}^*$ [see Fig.\ \ref{fig:Freqs_local}(a)]. Similarly at the inside of the horizon $q_1$ is the momentum of the gapless left-moving mode around $q=0$ while $q_2$ is the momentum of the right-moving mode around $q_{\text{in}}^*$  [see Fig.\ \ref{fig:Freqs_local}(b)]. 

Matching the current on both sides via \eqref{eq:cons_current} yields 
\begin{equation}
    |A_R|^2-|A_L|^2 = |B_R|^2-|B_L|^2,\label{eq:currentcons}
\end{equation}
where we used that $k_1=k_{out}^*-k_2$, $q_1=q_{in}^*-q_2$ and $k_2=k_{out}^*+q_2$ with $k^*_{out}=-q^*_{in}=2\arccos\left(\mu/\sqrt{\gamma^2+\mu^2}\right)$.

To obtain the scattering state corresponding to the scattering process in Fig.\ \ref{fig:Snapshot_local}, we take $A_L=0$ and $A_R=1$. Indeed, this situation corresponds to a plane wave in the left region $j < j_L$ scattering of the horizon, resulting in a transmitted and reflected part. We proceed by solving the Schrödinger-like equation numerically, starting from the initial values $f_\omega(j_R+1)=e^{ik_2(j_R+1)}$ and $f_\omega(j_R)=e^{ik_2j_R}$ and iteratively solving for $f_\omega(j-1)$ until finally arriving at $f_\omega(j_L-1)$. The coefficients $B_L$ and $B_R$ can then be calculated with
\begin{align}
    B_L &= \frac{f_\omega(j_L-1)-e^{-iq_2}f_\omega(j_L)}{e^{iq_1 j_L}(e^{-iq_1}-e^{-iq_2})} \\ B_R &= \frac{f_\omega(j_L-1)-e^{-iq_1}f_\omega(j_L)}{e^{iq_2 j_L}(e^{-iq_2}-e^{-iq_1})} 
\end{align}
Note that from the current conservation Eq.\ \eqref{eq:currentcons} we have:
\begin{equation}
     1 = |A_R|^2 = |B_R|^2-|B_L|^2\,,
\end{equation}
which provides a nice numerical cross-check. The transmission $T$ and reflection coefficient $R$ for the scattering process in Fig.\ \ref{fig:Snapshot_local} are given by
\begin{equation}
    T = \frac{|A_R|^2}{|B_R|^2}, \qquad R = \frac{|B_L|^2}{|B_R|^2}\label{eq:transmissioncoeff}
\end{equation}
The wave packets used in Sec.\ \ref{sec:local} are in effect appropriate linear combinations of $f_\omega$'s such that they are localized in space at $x_0$ (at $t=0$) and in energy around some value $\omega$.

Since the negative momentum modes around $q_{in}^*<0$ are occupied in the initial Minkowski ground state, we can infer the asymptotic occupation numbers $N_{out,\omega}$ for right moving wave packets on the outside and $N_{in,\omega}$ for left moving wave packets on the inside from the transmission and reflection coefficients:
\begin{equation}
    N_{out,\omega} =1-N_{in,\omega}= \frac{|A_R|^2}{|B_R|^2} = \frac{1}{|B_R|^2}\,,
\end{equation}
which numerically is found to coincide very well with the Dirac-Fermi spectrum [Eq.\ (\ref{eq:FD})], $N_{out,\omega} = f(\omega)$, $N_{in,\omega}=f(-\omega)$, with $T_H= \kappa /(2\pi k_B) = \hat{\kappa}\gamma /(2\pi k_B)$ in agreement with the previous sections.

\subsection{Analytical transmission coefficient by continuum approximation}\label{subsecD}
By mapping the lattice problem from the previous section onto a continuous scattering problem we can derive the observed Fermi-Dirac form of the transmission coefficient. The goal is to solve Eq. \eqref{eq:Schrodingerlike} and find the coefficient $B_R$ defined in Eq. \eqref{eq:planewaves}, from which the transmission coefficient readily follows. For this, we first perform a unitary transformation $c_n\rightarrow e^{i\alpha_n}c_n$, which changes the hopping coefficients of the Hamiltonian in Eq. \eqref{local}: $\mu+i\gamma_n\rightarrow \tau_n \in \mathbb{R}$. The recursion relation to solve is then
\begin{equation}
    \tau_n f_\omega(n+1) + \tau_{n-1}f_\omega(n-1) + (\mu+\omega)f_\omega(n) = 0\,,
    \label{eq:recurrence_real}
\end{equation}
with $\tau_n = \frac{1}{2}\sqrt{\mu^2+\gamma^2\tanh^2(\hat{\kappa} n)}$. As before, we are interested in solutions with the asymptotic form of Eq.\ \eqref{eq:planewaves}, with boundary conditions $A_R=1$ and $A_L=0$. Using a discrete Wentzel-Kramers-Brillouin (WKB) approximation \cite{Dingle,Geronimo,Braun}, the solutions of Eq. \eqref{eq:recurrence_real} can be written as exponentials with slowly varying phases. The quantity 
\begin{equation}   
    B(n) =  \frac{\mu+\omega}{2\sqrt{\tau_n \tau_{n-1}}} \approx \frac{\mu+\omega}{2\tau_n } = \frac{\mu+\omega}{\sqrt{\mu^2+\gamma^2 \tanh^2 \hat{\kappa}x} }\,,
\end{equation}
defines the turning points $\pm n_t$ via $B(n_t)=1$ and determines whether the phases in the WKB ansatz are real or complex. For $|n|> |n_t|$ the solutions are complex exponentials which asymptotically reduce to the plane waves in Eq.\ \eqref{eq:planewaves}. A crucial observation is, as explained in Appendix \ref{app:analytical_discr}, that in the vicinity of the turning points, we can write $f_\omega(n) = (-1)^n  P_\omega(n)/\sqrt{4\tau_n \tau_{n-1}}$, where $P_\omega(n)$ is slowly-varying and can be identified as the (WKB) solution of the following continuous Schr\"{o}dinger equation:
\begin{equation}   
    P_\omega''(x) +k(x)^2 P_\omega(x) = 0\,,
    \label{eq:Schrodinger}
\end{equation}
with $k(x)=\frac{1}{a}\sqrt{2(1-B(x))}$. The calculation of $B_R$ and $B_L$, which determine $N_{out,\omega}$, then amounts to finding the transmission coefficient $T_{\text{cont}}$ for a continuous scattering problem where a particle with mass $m=(\mu+\omega)/(\gamma a)^2$ and energy $E=\frac{\gamma^2}{\mu+\omega}$ scatters off a potential of the form
\begin{equation}
V(x)=\frac{\gamma^2}{\sqrt{\mu^2+\gamma^2 \tanh^2 \hat{\kappa}x/a}}\,.    
\end{equation}
For a quadratic potential barrier $V(x) = U_0 - c^2 x^2$, $T_{\text{cont}}$ is given by \cite{Berz,Kemble}
\begin{equation}
    T_{\text{cont}} = \frac{1}{1+e^K }\,,
\end{equation}
where $K = \pi\sqrt{2m} \left(\frac{U_0-E}{c}\right)$.
Approximating the potential by a parabola $V(x) \approx \gamma^2(\frac{1}{\mu} - \frac{\kappa^2}{2a^2\mu^3}x^2)$, and using the expressions for $m$ and $E$, we obtain 
\begin{equation}
    K = \frac{2\pi\mu}{\kappa} \left(1-\frac{1}{1+\frac{\omega}{\mu}} \right)\sqrt{1+\frac{\omega}{\mu}} \approx \frac{2\pi\omega}{\kappa}\left(1 + \mathcal{O}\left(\frac{\omega}{\mu}\right)\right)\,.
    \label{eq:parabolicresult}
\end{equation}
We thus conclude that for small $\omega$ we do indeed find a Fermi-Dirac distribution:
\begin{equation}
N_{out,\omega} = \frac{1}{1+e^{2\pi \omega/\kappa} }\,,
\end{equation}
which explains the numerical results of Secs.\ \ref{sec:local} and \ref{sec:scat}.

\begin{figure}
    \centering
    \includegraphics[width=0.75\linewidth]{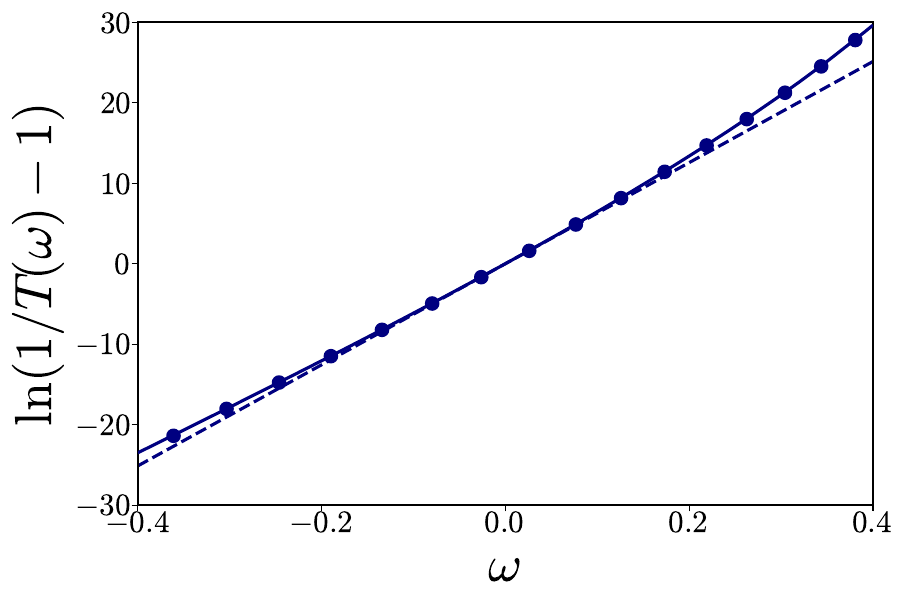}
    \caption{Comparison between the exponential factor $\ln (1/T-1)$ of the transmission coefficient $T$ from the numerical solution of \eqref{eq:app_recurrence} (dots) and the analytical prediction \eqref{eq:expfactor} (line) for $\hat{\kappa}=0.1$ and $\mu=0.5\gamma$. The leading order contribution $K=\frac{2\pi \omega}{\kappa}$ is shown with the dashed line.}
    \label{fig:expfactor}
\end{figure}

In Appendix \ref{app:analytical_discr}, we perform a more detailed calculation which uses the exact form of the potential and which also takes corrections to the continuum approximation into account. Here we simply give the result, which is
\begin{equation}\label{eq:expfactor}
  K = \frac{2\pi}{\hat{\kappa}}\left (\atan \frac{\mu+\omega}{\sqrt{\gamma^2+\mu^2-(\mu+\omega)^2}} -\atan \frac{\mu}{\gamma}\right)\,.
\end{equation}
The remarkable agreement between this analytical expression and the numerical solution is shown in Fig.\ \ref{fig:expfactor}. For small $\omega$, we can write
\begin{equation}
 \frac{\kappa}{2\pi\omega}K \approx 1 + \frac{\mu}{2\gamma} \left(\frac{\omega}{\gamma} \right) + \mathcal{O}\left(\frac{\omega^2}{\gamma^2}\right)\,,
\end{equation}
which shows that the leading order correction to the exponent in the Fermi-Dirac distribution is proportional to $\mu/2\gamma$. Notice that for the small temperatures $\kappa/(2\pi k_B)$ that we consider, this correction only appears at the tails of the Fermi-Dirac distribution.
\section{Conclusions}
We have presented two elementary free fermion lattice models which display the Hawking effect in quench dynamics: one Floquet model with a causal horizon, and one local Hamiltonian model without a causal horizon. We find that, for energies well below the lattice cut-off, both these models reproduce the thermal Hawking distribution found in continuum calculations. The main feature of the free fermion models is that they allow for a straightforward identification of the origin of the Hawking particles. 

These results now pave the way for exploring the Hawking effect in other quantum many-body systems, such as for example interacting spin chains (both analytically in integrable models, and numerically using, e.g., matrix product state simulations), 2D materials with tilted Dirac cones \cite{Jafari1,Jafari5,Jafari6}, and cold-atom or trapped ion experiments. Furthermore, the approach to Hawking radiation adopted in this work is closely related to previous studies of quantum quenches corresponding to moving defects \cite{Bastianello2018,BastianelloDeLuca2018,DeLuca2020}. It would be interesting to apply the analytical techniques developed for the moving defect problem, which also rely on solving a scattering problem in the comoving frame, to the Hawking quench protocol. Similarly, our work has a natural connection to the superluminal moving mass front which can be used to prepare ground states of gapless systems with emergent Lorentz invariance \cite{Agarwal2018,Mitra2019}. Our reformulation of the Hawking radiation phenomenon thus shows that it has many connections to previously studied topics in the field of many-body quantum dynamics, and we hope that it will inspire future work where techniques developed in the quantum dynamics community will be applied to study Hawking radiation in interacting systems. 
\vspace{0.25cm}
\begin{acknowledgments}
We thank Elliot Blommaert for an initial collaboration on the project and acknowledge inspiring discussions with Gertian Roose and Thomas Mertens. We would also like to thank an anonymous referee for their suggestion to analytically derive the transmission coefficient by a continuum mapping. This work has received funding from the European Research Council (ERC) under the European Union's Horizon 2020 research and innovation programme [Grant Agreements No. 715861 (ERQUAF)]. N.B. was supported by a postdoctoral research fellowship of the Research Foundation Flanders (FWO).

\end{acknowledgments}

\bibliography{bib}

\pagebreak
\begin{appendix}
\onecolumngrid

\section{Derivation of Hawking radiation in the continuum}\label{app:continuum}
In this appendix we review the derivation of Hawking radiation for a continuum Dirac fermion in one spatial dimension. We start by first defining the ``Minkowski" Hamiltonian, which is given by (working in units $\hbar$=1)

\begin{equation}
\hat{H}_0= \int \mathrm{d}x\, \psi^\dagger(x)\left( -i v \partial_x\right)\psi(x)\, .
\end{equation}
Here, $v$ is a velocity, and $\psi^\dagger(x)$, $\psi(x)$ are creation and annihilation operators satisfying the canonical fermionic anti-commutation relations:

\begin{eqnarray}
\{\psi^\dagger(x),\psi(x') \} & = & \delta(x-x') \\
\{\psi^\dagger(x),\psi^\dagger(x')\} & = & \{\psi(x),\psi(x')\} = 0 \, .
\end{eqnarray}
The Minkowski Hamiltonian $\hat{H}_0$ describes the right-moving part of a massless Dirac fermion in one spatial dimension. We ignore the left-moving part of the Dirac fermion because it does not play any role in the Hawking effect. The ground state $|\psi_0\rangle$ of $H_0$ is obtained by occupying all plane wave state with negative energy $E(k) = v k$, and thus satisfies

\begin{equation}
    \langle\psi_0|\psi^\dagger_k \psi_{k'}|\psi_0\rangle = \Theta(-k)\delta(k-k')\,,
\end{equation}
with $\psi^\dagger_k = (2\pi)^{-1/2} \int \mathrm{d}x\, e^{ikx} \psi^\dagger(x)$. 
\\ \\
Besides the Minkowski Hamiltonian, we also need a horizon Hamiltonian $\hat{H}_h$, which is given by

\begin{equation}\label{Hhor}
\hat{H}_h = \int\mathrm{d}x\,\psi^\dagger(x)\left(-i v(x) \partial_x -\frac{i}{2}\partial_x v(x)\right)\psi(x) := \int\mathrm{d}x\,\psi^\dagger(x)\hat{h}_h\psi(x)\, .
\end{equation}
In the horizon Hamiltonian, the velocity $v(x)$\footnote{Note that here we use $v(x)$ for the full velocity of the right-moving field. For instance, for an acoustic metric of the form $ds^2=(c(x)^2-v_F(x)^2)dt^2-2 v_F(x) dx dt -dx^2$ we would have $v(x)=c(x)-v_F(x)$.} is a continuous function of $x$ which satisfies $v(x)>0$ for $x>0$, and $v(x)<0$ for $x<0$. The horizon is located at $x=0$, where $v(0)=0$. The region $x>0$ is called the `outside' region of the horizon, and $x<0$ is the `inside' region. In the presence of an additional mode which is everywhere left moving (recall that we ignoring this mode here), it is possible to move both left and right for $x>0$, whereas for $x<0$ one is forced to move to the left which means that it is impossible to cross the horizon (one is trapped behind the horizon). The term $-i\partial_x v(x) \psi^\dagger(x)\psi(x)/2$ is necessary to ensure that the horizon Hamiltonian is Hermitian.
\\ \\
In this appendix we will be interested in two different velocity profiles:

\begin{eqnarray}
v(x) & = & \kappa x \,,\hspace{2.08 cm} \left[\text{Unruh case}\right]\\
v(x) & = & v \tanh(x\kappa/v)\, . \hspace{0.5 cm}\left[\text{Hawking case}\right]
\end{eqnarray}
Here, $\kappa$, which has dimensions of inverse time (or energy since $\hbar = 1$), is the `surface gravity'. The first velocity profile $v(x) = \kappa x$ corresponds to a Rindler observer, which has a constant acceleration. This case was analysed by Unruh, and we will therefore call this the Unruh case. The second velocity profile $v(x)=v\tanh(\kappa x/v)$ goes to $\pm v$ as $x\rightarrow \pm \infty$. So in this case, the horizon Hamiltonian is a `black hole' Hamiltonian which matches the Minkowski Hamiltonian far away on the right hand side of the horizon (this is the `outside' of the black hole). The black hole case was analysed by Hawking, and so we refer to it as the Hawking case. Below, we discuss the Unruh and Hawking cases separately. 

\subsection{Unruh/Rindler case}

We start by finding the eigenstates of the single-particle Hamiltonian $\hat{h}_h$ in Eq.\ \eqref{Hhor} with the Unruh velocity profile $v(x) = \kappa x$. For this we use the ansatz $\varphi(x) = \exp(iS(x))$. This state is an eigenstate with energy $\omega$ if

\begin{equation}
\omega = v(x)\partial_x S(x) - \frac{i}{2}\partial_x v(x)
\end{equation}
The solutions to this equation are

\begin{eqnarray}
S^{\pm}(x) & = & \lim_{\epsilon \rightarrow 0^\pm}\int^x \mathrm{d}x'\,\frac{\omega + i \partial_{x'} v(x')/2}{v(x')  + i\epsilon } \label{eps}\\ 
& = & \mathcal{P} \int^x \mathrm{d}x'\, \frac{\omega}{v(x')} \mp i \frac{\pi\omega}{ \kappa}\Theta(x)\\ 
 & & +\frac{i}{2}\mathcal{P}\int^x \mathrm{d}x'\, \partial_{x'}\ln(v(x')) \pm \frac{\pi}{2}\Theta(x) \\
  & = & \mathcal{P} \int^x \mathrm{d}x'\, \frac{\omega}{v(x')}\mp i \frac{\pi\omega}{ \kappa}\Theta(x)  +\frac{i}{2}\ln(|v(x)|) \pm \frac{\pi}{2}\Theta(x)\, ,
\end{eqnarray}
where $\mathcal{P}$ denotes the Cauchy principal value and $\Theta(x)$ is the Heaviside step function. A similar $i\epsilon$ prescription was also previously used in Refs. \cite{DamourRuffini,Srinivasan,Wilczek}. Plugging these solutions for $S(x)$ back in the ansatz $\varphi(x) = \exp(iS(x))$, we find two solutions for the Rindler eigenstates with energy $\omega$:

\begin{eqnarray}
\varphi_\omega^{R+}(x) & = & \Theta(-x)\frac{1}{\sqrt{-\kappa x}}e^{i\omega \ln(-\kappa x)/\kappa} + \Theta(x)\frac{e^{i\pi/2}e^{\pi\omega/\kappa}}{\sqrt{\kappa x}}e^{i\omega \ln(\kappa x)/\kappa} \label{Rindplus}\\
\varphi_\omega^{R-}(x) & = & \Theta(-x)\frac{1}{\sqrt{-\kappa x}}e^{i\omega \ln(-\kappa x)/\kappa} + \Theta(x)\frac{e^{-i\pi/2}e^{-\pi\omega/\kappa}}{\sqrt{\kappa x}}e^{i\omega \ln(\kappa x)/\kappa}\label{Rindmin}
\end{eqnarray}
Using 

\begin{eqnarray}
    \int_0^\infty\mathrm{d}x\, \frac{1}{\kappa x} e^{i(\omega-\omega') \ln(\kappa x)/\kappa} & = & \int_{-\infty}^\infty \mathrm{d}(\ln(\kappa x)/\kappa) e^{i(\omega-\omega') \ln(\kappa x)/\kappa} \\
    & = & 2\pi \delta(\omega-\omega')
\end{eqnarray}
we see that the Rindler wavefunctions in Eqs. \eqref{Rindplus}-\eqref{Rindmin} satisfy

\begin{equation}
    \int\mathrm{d}x\, \left[\varphi^{Rs}_\omega(x)\right]^*\varphi^{Rs'}_{\omega'}(x) = 2\pi (1+e^{s2\pi \omega/\kappa})\delta_{s,s'}\delta(\omega-\omega')\, ,
\end{equation}
where $s,s' = \pm$.
\\ \\
A crucial observation is that the Rindler eigenfunctions can be written as

\begin{eqnarray}
\varphi_{\omega}^{R+}(x) & = & \int_0^\infty \mathrm{d}k\, e^{ikx} \tilde{\varphi}^{R+}_{\omega}(k) \, ,\\
\varphi_{\omega}^{R-}(x) & = & \int_0^\infty \mathrm{d}k\, e^{-ikx} \tilde{\varphi}^{R-}_{\omega}(k)\, .
\end{eqnarray}
As pointed out by Unruh \cite{Unruh1976}, this follows from the fact that any function which only has positive (negative) Fourier modes is analytic in the upper (lower) half of the complex plane. The branch cuts of the logarithm and the square root in $\varphi^{R+}_{\omega}(x)$ ($\varphi^{R-}_{\omega}(x)$) both lie in the lower (upper) half-plane, which means that $\varphi^{R+}_{\omega}(x)$ ($\varphi^{R-}_{\omega}(x)$) is analytic in the upper (lower)  half plane. From this observation we learn that $\varphi^{R+}_{\omega}(x)$ is made entirely from states which are unoccupied in the Minkowski vacuum, whereas $\varphi^{R-}_{\omega}(x)$ made entirely from states which are occupied in the Minkowski vacuum. Because the set of occupied (unoccupied) plane wave states in the Minkowski vacuum and the Rindler wavefunctions $\varphi^{R-}_\omega(x)$ ($\varphi^{R+}_\omega(x)$) are related via an invertible linear transformation, we conclude that

\begin{equation}\label{Mink}
\langle \psi_0|\psi^\dagger_{R,s}(\omega)\psi_{R,s'}(\omega')|\psi_0\rangle = \begin{cases}
      \delta(\omega-\omega') & \text{ if }s = s' = -\\
      0 & \text{otherwise}
    \end{cases}\, ,
\end{equation}
where

\begin{eqnarray}
 \psi_{R-}^\dagger(\omega) & =  & \frac{1}{\sqrt{1+e^{-2\pi\omega/\kappa}}}\frac{1}{\sqrt{2\pi}}\int\mathrm{d}x\, \varphi^{R-}_{\omega}(x)\psi^\dagger(x)\,, \\
 \psi_{R+}^\dagger(\omega) & = & \frac{1}{\sqrt{1+e^{2\pi\omega/\kappa}}}\frac{1}{\sqrt{2\pi}}\int\mathrm{d}x\, \varphi^{R+}_{\omega}(x)\psi^\dagger(x)\, ,
\end{eqnarray}
are fermion creation operators satisfying

\begin{eqnarray}
\{\psi^\dagger_{Rs}(\omega),\psi_{Rs'}(\omega')\} & = & \delta_{s,s'} \delta(\omega-\omega') \\
\{\psi^\dagger_{Rs}(\omega),\psi^\dagger_{Rs'}(\omega')\} & = & \{\psi_{Rs}(\omega),\psi_{Rs'}(\omega')\} = 0\, .
\end{eqnarray}
Finally, we define the `single-sided' eigenstate creation operators as follows:

\begin{eqnarray}
\psi^\dagger_{R,out}(\omega) & = & \frac{1}{\sqrt{2\pi}}\int_0^\infty \mathrm{d}x\, \frac{1}{\sqrt{\kappa x}}e^{i\omega 
\ln(\kappa x))/\kappa } \psi^\dagger(x),\\
\psi^\dagger_{R,in}(\omega) & = & \frac{1}{\sqrt{2\pi}}\int_{-\infty}^0 \mathrm{d}x\, \frac{1}{\sqrt{-\kappa x}} e^{i\omega \ln(-\kappa x)/\kappa } \psi^\dagger(x)\, .
\end{eqnarray}
These creation operators again satisfy the canonical fermionic anti commutation relations:

\begin{eqnarray}
\{\psi^\dagger_{R,l}(\omega),\psi_{R,l'}(\omega')\} & = & \delta_{l,l'} \delta(\omega-\omega')\, , \\
\{\psi^\dagger_{R,l}(\omega),\psi^\dagger_{R,l'}(\omega')\} & = & \{\psi_{R,l}(\omega),\psi_{R,l'}(\omega')\} = 0\, ,
\end{eqnarray}
where $l,l' = in/out$. From their definition, it is clear that the single-sided eigenmodes can be obtained as linear combinations of the degenerate wavefunctions $\varphi^{R+}_\omega(x)$ and $\varphi^{R-}_\omega(x)$. In particular, it holds that

\begin{eqnarray}
\psi^\dagger_{R,in}(\omega) & = & \frac{1}{\sqrt{1+e^{2\pi \omega/\kappa}}} \psi^\dagger_{R,+}(\omega) + \frac{1}{\sqrt{1+e^{-2\pi \omega/\kappa}}} \psi^\dagger_{R,-}(\omega)\, ,\nonumber \\
\psi^\dagger_{R,out}(\omega) & = & -i\frac{e^{\pi\omega/\kappa}}{\sqrt{1+e^{2\pi \omega/\kappa}}} \psi^\dagger_{R,+}(\omega) +i \frac{e^{-\pi\omega/\kappa}}{\sqrt{1+e^{-2\pi \omega/\kappa}}} \psi^\dagger_{R,-}(\omega).\label{RindlertoUnruh}
\end{eqnarray}
From these relations, and Eq.\ \eqref{Mink}, we find that 

\begin{equation}
\langle \psi_0|\psi^\dagger_{R,out}(\omega)\psi_{R,out}(\omega')|\psi_0\rangle = \frac{1}{e^{2\pi \omega/\kappa}+1} \delta(\omega-\omega'):= f(\omega) \delta(\omega-\omega').\label{unruh}
\end{equation}
This is the famous Unruh result \cite{Unruh1976}: after tracing out the region behind the horizon of an accelerating Rindler observer, the density matrix of the outer region is a thermal state with temperature $T = \kappa/2\pi k_B$ of the Rindler modes which only have support on the outside of the horizon.  
The Bogoliubov transformations \eqref{RindlertoUnruh} can also be used to compute the occupation numbers for the inside (left-moving) modes and the correlation between the inside and outside modes:
\bea
\langle \psi_0|\psi^\dagger_{R,in}(\omega)\psi_{R,in}(\omega')|\psi_0\rangle &=& \frac{1}{e^{-2\pi \omega/\kappa}+1} \delta(\omega-\omega')= f(-\omega)\delta(\omega-\omega')\,,\label{insideocc}\\
\langle \psi_0|\psi^\dagger_{R,in}(\omega)\psi_{R,out}(\omega')|\psi_0\rangle &=& -i\frac{e^{-\pi\omega/\kappa}}{e^{-2\pi \omega/\kappa}+1} \delta(\omega-\omega')=-i\sqrt{f(\omega)f(-\omega)}\delta(\omega-\omega')\,,\label{cross}
\eea
Note that we have also found that the Minkowski ground state is an eigenstate of the Rindler Hamiltonian: $H_h=\int \!d\omega\, \omega\,\left( \psi^\dagger_{R+}(\omega)\psi_{R+}(\omega)+\psi^\dagger_{R-}(\omega)\psi_{R-}(\omega)\right)$ . So if we were to quench the Minkowski ground state with the Rindler Hamiltonian, this would not generate any non-trivial dynamics.

\subsection{Hawking/black hole case}

\begin{figure}
    \centering
    \includegraphics[width=0.85\linewidth]{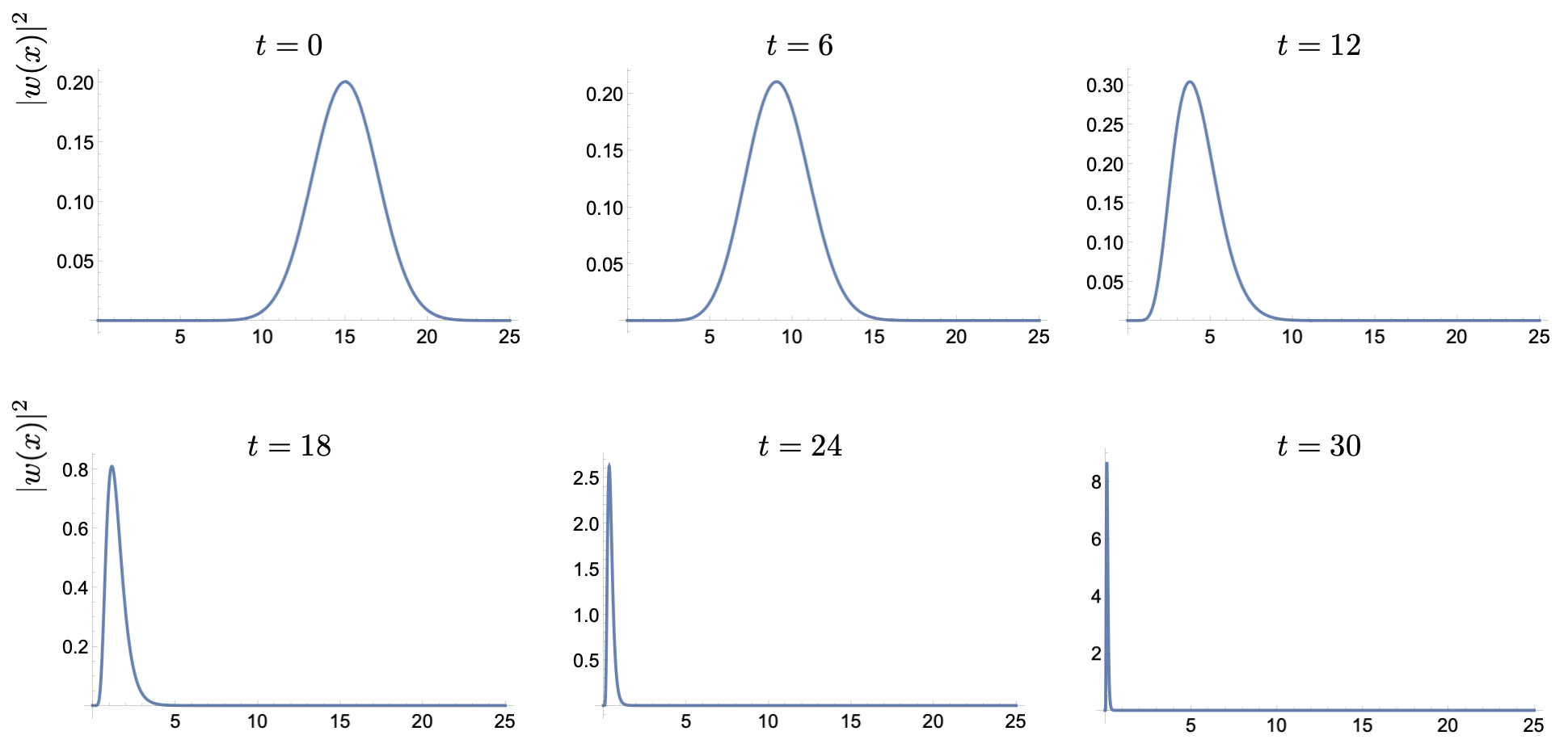}
    \caption{Backwards-in-time evolution of the wave packet $w_{x_0,k_0}(x,-t)$ (Eq.\ \eqref{w(t)}) with $x_0 =15,k_0=5,\sigma=0.25,v=1$ and $\kappa=1/5$. Notice the alteration in the wave packet behaviour upon entering the quantum atmosphere at $x\approx v/\kappa=5$. }
    \label{fig:QFTwavepacket}
\end{figure}

Let us now consider the horizon Hamiltonian in Eq.\ \eqref{Hhor} with the black hole velocity profile $v(x) = v \tanh(\kappa x/v)$. We want to show that Hawking radiation emerges upon quenching the Minkowski ground state with this Hamiltonian. 
\\ \\
For large positive $x$, the black hole velocity profile approaches a constant velocity $v$. So far outside the black hole, the Minkowski ground state looks like the ground state of the black hole Hamiltonian, and no non-trivial dynamics will originate from this region. For large negative $x$, the black hole velocity profile approaches the constant velocity $-v$. In this region, the Minkowski ground state is again an eigenstate of the black hole Hamiltonian, but it is no longer the ground state. Instead, $|\psi_0\rangle$ locally looks like the highest energy eigenstate of $\hat{H}_h$. Nevertheless, the region far away on the inside of the black hole will again not generate any non-trivial dynamics. Close to the horizon, we can approximate the black hole velocity profile as $v(x) \sim \kappa x$, which is exactly the Rindler velocity profile. From our previous analysis of the Unruh/Rindler case, we conclude that close to the horizon, the Minkowski ground state looks like a finite-temperature state with temperature $T=\kappa/2\pi k_B$ for the black hole Hamiltonian. Again we expect that the region close to the horizon where $v(x)\sim \kappa x$ will not generate any non-trivial dynamics, because in the previous section, we found that the Minkowski ground state is an eigenstate of the Rindler Hamiltonian.
\\ \\
The previous discussion suggests the following physical picture: for the black hole Hamiltonian, the Minkowski ground state looks like a state with a temperature profile $T(x) \propto \partial_x |v(x)|$. Far on the outside, the temperature is $T(x\gg v\kappa^{-1}) \sim 0^+$ and the Minkowski ground state is effectively the black hole ground state. Far on the inside, $T(x\ll -v\kappa^{-1}) \sim 0^-$, and the Minkowski ground state looks like the highest energy state. Close to the horizon on the outside, $T(0<x\ll v\kappa^{-1}) = \kappa/2\pi k_B$. And close to the horizon on the inside, $T(0> x \gg -v\kappa^{-1}) \sim -\kappa/2\pi k_B$ [see Eq.\ \eqref{insideocc}]. Gradients in the temperature occur in regions where the velocity has a non zero curvature: $\partial_x T(x) \propto \partial_x^2 v(x)$. The Hawking effect corresponds to the emission of particles from the ``hot'' to the ``cold'' regions in the Minkowski ground state under a quench with the black hole Hamiltonian. These particles originate from the regions where $\partial_x^2v(x) \neq 0$, which is roughly at a distance $v\kappa^{-1}$ away from the horizon.
\\ \\
To make this physical picture more concrete, we will first calculate the occupation number of wave packets on the outside of the horizon. We define the wave packet creation operator as: \be W^\dagger_{x_0, k_0}=\int\! dx\, w_{x_0,k_0} (x) \psi^\dagger(x)\,, \ee
with \be w_{x_0,k_0}(x)=\frac{1}{(2\pi\sigma^2)^{1/4}}\int^{+\infty}_{-\infty}\! dk \, e^{-\frac{(k-k_0)^2}{4\sigma^2}}\frac{1}{\sqrt{2\pi}} e^{ik (x-x_0)}\,,\label{wavepacket}\ee
a Gaussian wave-packet, with the momentum centered around $k_0$ and the position centered around $x_0$. The occupation number  $N_{x_0,k_0}(t)$ of this packet can be found as:
\bea N_{x_0,k_0}(t)&=&\bra{\psi_0(t)} W^\dagger_{x_0, k_0} W_{x_0, k_0}\ket{\psi_0(t)}\nonumber\\ &=&\bra{\psi_0} e^{i t H_h} W^\dagger_{x_0, k_0} W_{x_0, k_0} e^{-i t H_h}\ket{\psi_0}\nonumber\\
&=&\bra{\psi_0}  W^\dagger_{x_0, k_0}(-t) W_{x_0, k_0}(-t)\ket{\psi_0}\,,\eea
with \bea W^\dagger_{x_0, k_0}(-t)&:=&\int\! dx\, w_{x_0,k_0}e^{i t H_h}  \psi^\dagger(x) e^{-i t H_h} \nonumber\\
&=&\int\! dx\, w_{x_0,k_0}(x) e^{+i t h_h^*} \psi^\dagger(x)\,\nonumber\\
&=&\int\! dx\, \left(e^{+i t h_h} w_{x_0,k_0}(x)\right) \psi^\dagger(x)\nonumber\\ 
&=&\int\! dx\, w_{x_0,k_0}(x,-t) \psi^\dagger(x)\,,\eea
where in the last line the wave-packet $w_{x_0,k_0}(x,-t)=e^{+i t h_h} w_{x_0,k_0}(x)$ is propagated back(!) in time with the single-particle Hamiltonian $h_h$ (cfr. Eq.\ \eqref{Hhor}). To obtain $w_{x_0,k_0}(x,-t)$, it is useful to look at the expressions for the eigen out-modes, which from a generalization of the Unruh/Rindler case, are readily found as: \be \varphi_\omega^{out}(x)=\Theta(x)\frac{1}{\sqrt{2\pi v(x)}} e^{i\omega \int^x d x'\, \frac{1}{v(x')} }=\Theta(x)\frac{1}{\sqrt{2\pi v\tanh(\kappa x/v)}}e^{i\omega \ln(\sinh(\kappa x/v))/\kappa}\,. \ee
Far away from the horizon, $x\gg v/\kappa$, we effectively have Minkowski space-time and $\varphi_\omega^{out}$ reduces to a plane wave; on the other hand, close to the horizon, $x\ll v/\kappa$, we have effectively the Rindler case and $\varphi_\omega^{out}$ reduces to $\varphi_\omega^{R,\, out}$, explicitly:
\be \varphi^{out}_\omega(x)\stackrel{x\gg v/\kappa}{=}\frac{1}{\sqrt{2\pi v}} e^{i\frac{\omega}{v}(x-\frac{v}{\kappa}\ln(2))}\quad\quad\varphi^{out}_\omega(x)\stackrel{x\ll v/\kappa}{=}\frac{1}{\sqrt{2\pi \kappa x}} e^{i\omega \ln(\kappa x/v)/\kappa }= e^{-i\omega \ln(v)/\kappa}\varphi_\omega^{R, out}(x)\,. \label{phibehavior}\ee
Taking the initial wave-packet $w_{x_0,k_0}$ (Eq.\ \ref{wavepacket}) far from the horizon, $x_0\gg v/\kappa$, we can now effectively write: 
\be w_{x_0,k_0}(x)= \frac{1}{(2\pi\sigma^2 v^2)^{1/4}}\int^{+\infty}_{-\infty}\! d\omega \, e^{-\frac{(\omega-vk_0)^2}{4\sigma^2v^2}}e^{i\frac{\omega}{v}(\frac{v}{\kappa}\ln(2)-x_0)}\varphi_\omega^{out}(x)\,,\ee
which in turn allows to find the time-evolved wave-packet:

\bea w_{x_0,k_0}(x,-t)&=&  \frac{1}{(2\pi\sigma^2 v^2)^{1/4}}\int^{+\infty}_{-\infty}\! d\omega \, e^{-\frac{(\omega-vk_0)^2}{4\sigma^2v^2}}e^{i\frac{\omega}{v}(\frac{v}{\kappa}\ln(2)-x_0+v t)}\varphi_\omega^{out}(x)\,,\label{w(t)}\\
&=&\frac{1}{\sqrt{v(x)}}  \frac{1}{(2\pi\sigma^2 v^2)^{1/4}}\int^{+\infty}_{-\infty}\! d\omega \, e^{-\frac{(\omega-vk_0)^2}{4\sigma^2v^2}}e^{i\frac{\omega}{v}(\frac{v}{\kappa}\ln(2)-x_0+v t+y(x))}\,,\eea
where on the last line we have introduced the `tortoise'-coordinate $y(x)=\int^x dx' v/v(x')=v/\kappa\ln(\sinh(\kappa x/v)) $. Notice that \be y(x)\stackrel{x\gg v/\kappa}{=} x-\frac{v}{\kappa} \ln(2)
\quad\quad \quad y(x)\stackrel{x\ll v/\kappa}{=} \frac{v}{\kappa} \ln (\kappa x/v)\,,\ee
in particular we have that $y(x)\rightarrow -\infty$ when $x\rightarrow 0+$. We see that up to the pre-factor $v(x)^{-1/2}$ the wave-packet $w_{x_0,k_0}(x,-t)$ consists of plane-waves in $y$-space and therefore find that the wave-packet moves with constant width $\Delta y \sim 1/\sigma$ and constant velocity $-v$ in $y$-space: $y_c(t)=x_0 -\frac{v}{\kappa}\ln(2)- v t$, where $y_c(t)$ denotes the center-position of the packet. In $x$-space this translates to a wave-packet propagating with constant width and constant velocity $-v$, as long as $x_c(t)\gg v/\kappa$. This behavior changes when the wave-packet enters the `quantum-atmosphere'  $|x|\lesssim v/\kappa$, around $t=t^*\sim x_0/v-1/\kappa$. Inside the quantum atmosphere, the wave-packet will start to contract, $\Delta x\sim\Delta y/y'(x_c)\sim v(x_c)/(v\sigma)$ and slow down, $\dot x_c\sim\dot y_c/y'(x_c)= v(x_c)$, asymptotically ($t\rightarrow \infty$) `freezing' on the horizon $x=0$. We illustrate this behavior in Fig.\ \ref{fig:QFTwavepacket}.

Summarising, the wave-packet evolution has essentially two regimes: for $t\lesssim t^*$ it has only non-trivial support outside the quantum atmosphere, $x>v/\kappa$, while for $t\gtrsim t^*$ it will end up lying completely inside the quantum atmosphere $0<x<v/\kappa$. From \eqref{phibehavior} and \eqref{w(t)} we see that this implies, that for $t\lesssim t^*$ the wave-packet remains a superposition of Minkowski plane-waves, while for $t\gtrsim t^*$ it reduces to a superposition of Rindler eigen-modes, explicitly:
\bea W^\dagger_{x_0,k_0}(-t)&\stackrel {t\ll t^*}{=}& \frac{1}{(2\pi\sigma^2 v^2)^{1/4}}\int^{+\infty}_{-\infty}\! d\omega \, e^{-\frac{(\omega-vk_0)^2}{4\sigma^2v^2}}e^{i\frac{\omega}{v}(\frac{v}{\kappa}\ln(2)-x_0+v t)}\psi^\dagger_M(\omega)\\
W^\dagger_{x_0,k_0}(-t)&\stackrel {t\gg t^*}{=}& \frac{1}{(2\pi\sigma^2 v^2)^{1/4}}\int^{+\infty}_{-\infty}\! d\omega \, e^{-\frac{(\omega-vk_0)^2}{4\sigma^2v^2}}e^{i\frac{\omega}{v}(\frac{v}{\kappa}\ln(2)-x_0+v t)}\psi^\dagger_{R,out}(\omega)\eea
Finally, evaluating $N_{x_0,k_0}(t)=\bra{\psi_0}W^\dagger_{x_0,k_0}(-t) W_{x_0,k_0}(-t)\ket{\psi_0}$ we find that this gives a smeared out step-function $\Theta(-\omega)$ (with $\omega=k_0 v$) at initial times $t\ll t^*$; going over into the Fermi-Dirac distribution $f(\omega)$ at later times $t\gg t^*$ (see \eqref{unruh}), provided that $v\sigma \ll k_BT_H= \frac{\kappa}{2\pi}$. This is also precisely what we find in our lattice simulations as discussed in the main text. 

In a completely similar fashion one can consider the time evolution of a wave-packet, initially deep inside the black hole ($x_0\ll -v/\kappa$), resulting in (with now $t^*=-x_0/v-1/\kappa$):
\bea W^\dagger_{x_0,k_0}(-t)&\stackrel {t\ll t^*}{=}& \frac{1}{(2\pi\sigma^2 v^2)^{1/4}}\int^{+\infty}_{-\infty}\! d\omega \, e^{-\frac{(\omega+vk_0)^2}{4\sigma^2v^2}}e^{i\frac{\omega}{v}(\frac{v}{\kappa}\ln(2)+x_0+v t)}\psi^\dagger_M(-\omega)\\
W^\dagger_{x_0,k_0}(-t)&\stackrel {t\gg t^*}{=}& \frac{1}{(2\pi\sigma^2 v^2)^{1/4}}\int^{+\infty}_{-\infty}\! d\omega \, e^{-\frac{(\omega+vk_0)^2}{4\sigma^2v^2}}e^{i\frac{\omega}{v}(\frac{v}{\kappa}\ln(2)+x_0+v t)}\psi^\dagger_{R,in}(\omega)\,.\eea
This then gives $N_{x_0,k_0}(t)$ going from the smeared out step-function $\Theta(\omega)$ (with now $\omega=-k_0 v$) at $t\ll t^*$ to $f(-\omega)$ at times $t\gg t^*$, again in agreement with our lattice results. Finally, for the wave-packet correlation function (for $t\gg -x_{in}/v -1/\kappa, x_{out}/v-1/\kappa$ and with $k_{in}=-\omega_0/v,\,k_{out}=+\omega_0/v$) we get (from \eqref{cross}):
\bea\bra{\psi_0}W^\dagger_{x_{in}, k_{in}}(-t) W_{x_{out}, k_{out}}(-t)\ket{\psi_0}&=& -\frac{i}{(2\pi\sigma^2 v^2)^{1/2}}\int^{+\infty}_{-\infty}\! d\omega e^{-\frac{(\omega-\omega_0)^2}{2\sigma^2v^2}}e^{i\frac{\omega}{v}(x_{in}+x_{out})} \sqrt{f(\omega)f(-\omega)}\, \\
&\approx& -i \sqrt{f(\omega_0)f(-\omega_0)} e^{i\frac{\omega_0}{v}(x_{in}+x_{out})} e^{-\frac{\sigma^2(x_{in}+x_{out})^2}{2}}\,,\eea
where on the last line we again assumed $v\sigma\ll k_BT_H  $. This is the continuum version of our lattice results on the particle correlations (see e.g. Fig.\ \ref{fig:FLresults}(b) and the corresponding discussion in the main text).  

The region inside the quantum atmosphere $-\kappa/v\lesssim x \lesssim +\kappa/v$ is not altered during the evolution, yet it acts like an infinite reservoir for the outgoing radiation, with arbitrary short (`'transPlanckian'') wavelength modes, that upon leaving the horizon get redshifted to their asymptotic value (see, e.g., Fig.\ \ref{fig:QFTwavepacket}). This is very different for the lattice models that we discuss in the main text, where the outgoing radiation finds it origin in a bulk reservoir of ingoing doubler modes on the outside (for the Floquet model) or in a bulk reservoir of outgoing doubler modes on the inside of the boundary (for the local model).

\section{Details of numerical simulations for the Floquet model}\label{app:Floquet}

For our numerical calculations, we use the following explicit expression for $\gamma(x)$:

\begin{equation}\label{tx}
\gamma(x) = \gamma\left(\frac{4}{\pi b}\arctan\bigg[\frac{\cosh(\tilde{\kappa} \pi W/4)}{\cosh(\pi \tilde{\kappa}(x - L/2)/2)}\bigg] + \frac{b-1}{b} \right)^b \, ,
\end{equation}
where $L = Na$ is the length of the system, and $\tilde{\kappa}$ and $b$ are free parameters. We take the overall energy scale to be $\gamma = (\Delta t)^{-1}$, such that $a\gamma(x) = a/\Delta t = v_{Fl}$ at the location of the black hole horizon $x_b = L/2-W/2$, and at the location of the white hole horizon $x_w = L/2 + W/2$. The surface gravity is given by $\kappa = a\gamma\tilde{\kappa}\tanh( \tilde{\kappa}\pi W/4)$. For a sufficiently large width $W$ of the region outside the horizon (we always take $W$ to be a couple of hundred lattice sites), we effectively have $\kappa = \tilde{\kappa} v_{Fl}$. 

The wave-packet operators $W^\dagger_{x_0,\omega}$ are defined as follows:

\begin{eqnarray}
    W^\dagger_{x_0,\omega} & = & \mathcal{N}\sum_{k = -\pi/a}^{\pi/a} e^{-(k-\omega/v_{out})^2/4\sigma^2} e^{-i k x_0} c^\dagger_k \hspace{0.38 cm}\text{ for } x_b < x_0 <x_w\, ,\label{eq:wavepacket}\\
    W^\dagger_{x_0,\omega} & = & \mathcal{N}\sum_{k = -\pi/a}^{\pi/a} e^{-(k+\omega/v_{in})^2/4\sigma^2} e^{-ik x_0} c^\dagger_k \hspace{0.5 cm} \text{ for } x_0 < x_b\, ,
\end{eqnarray}
where $\mathcal{N}$ is a normalization factor, and $c^\dagger_k = \frac{1}{\sqrt{N}}\sum_{j=1}^{N}e^{ikaj}c^\dagger_j$. As mentioned in the main text, we always take $|\omega \Delta t| \ll 1$. The quantities $v_{out} = v(x\gg x_b)-v_{Fl}$ and $v_{in} = v_{Fl}-v(x\ll x_b)$ are (the absolute values of) the velocities of the wave packets on respectively the outside and inside of the black hole horizon. The velocity profile $v(x)$ is approximately constant for $|x-x_b|,|x-x_w|\gg \tilde{\kappa}^{-1}$.

\begin{figure}
    \centering
    \includegraphics[scale=0.4]{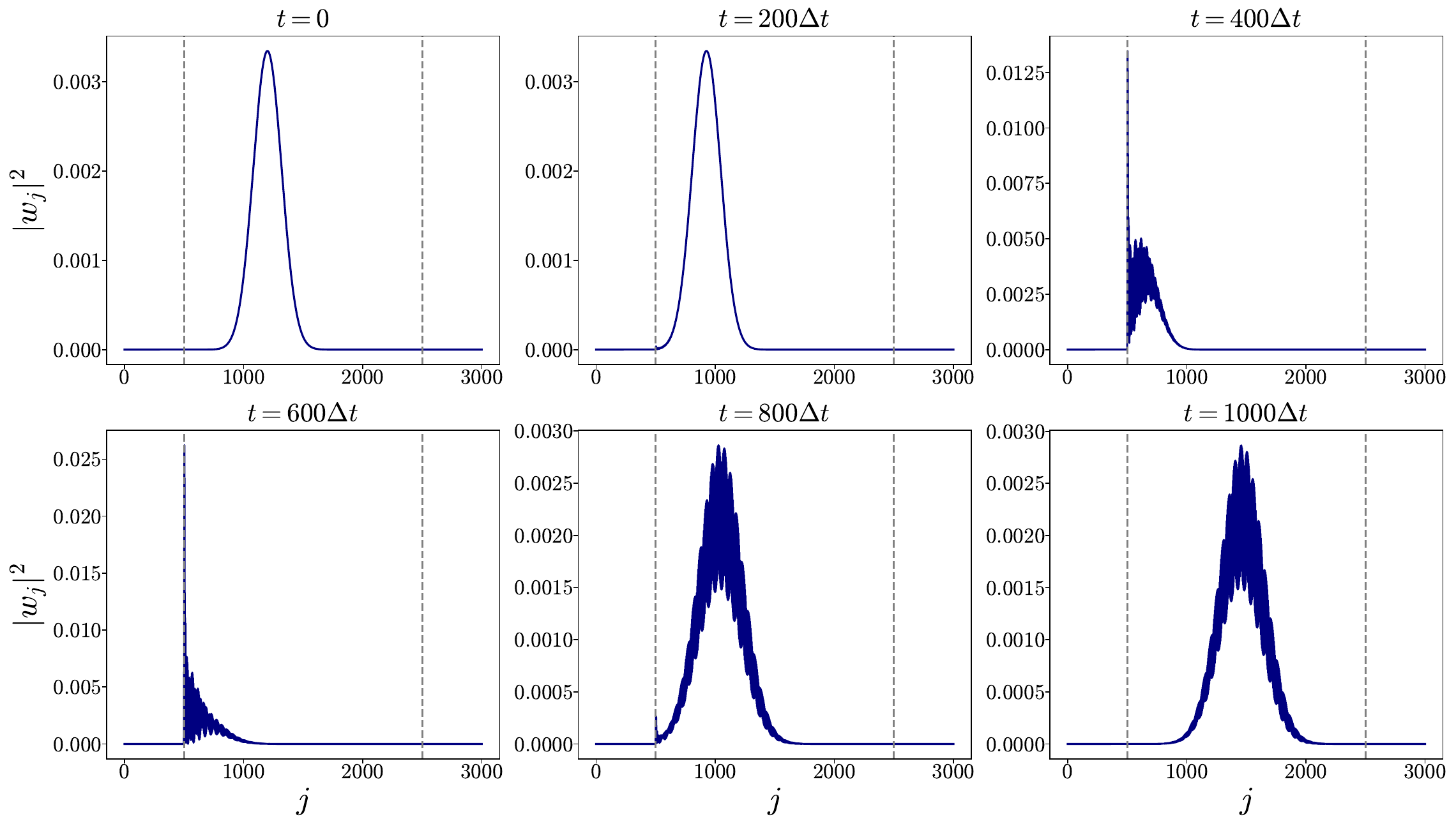}
    \caption{Backwards-in-time evolution of the wave packet $W^\dagger_{x_0,\omega} = \sum_j w_j c^\dagger_j$ with $x_0 = x_b + 700 a$ and $\omega = 0.057 t$. The width of the wave packet in momentum space is $\sigma = 2(2\pi/L)$. The system size is $N=3000$, $x_b = 500 a$ and $x_w = 2500 a$. The function $t(x)$ in Eq.\ \eqref{tx} was used with $\tilde{\kappa}a=0.1$ and $b=3$. }
    \label{fig:evolution}
\end{figure}

We calculate the time-dependent wave packet occupation number $N_{x_0,\omega}(n\Delta t)$ in practice by writing it as $N_{x_0,\omega}(n\Delta t) = \langle \psi_0|W^\dagger_{x_0,\omega}(-n\Delta t) W_{x_0,\omega}(-n\Delta t)|\psi_0\rangle$, where $W^\dagger_{x_0,\omega}(-n\Delta t) = U^{-n}(\Delta t)W^\dagger_{x_0,\omega}U^n(\Delta t)$ is the wave packet operator evolved \emph{backwards} in time; and where in analogy with the continuum case we define $W^\dagger_{x_0,\omega}(t):=\sum_i w_i(x_0,\omega,t) c_i^\dagger$, with $w_i(x_0,\omega,t)$ obtained from evolving with the single-particle Schr\"{o}dinger equation. In Fig.\ \ref{fig:evolution}, we show an example of such a backwards-in-time evolved wave packet on the outside of the black hole horizon. Note that because we evolve backwards in time, the right-moving wave packet moves to the left, i.e., it moves closer to the black hole horizon. From Fig.\ \ref{fig:evolution}, we see that during the time evolution the wave packet does not cross the horizon, but instead develops very short-wavelength oscillations and subsequently bounces back to the outside region. In Ref. \cite{CorleyJacobson}, this was interpreted as a Bloch oscillation.

\begin{figure}
    \centering
    \includegraphics[scale=0.5]{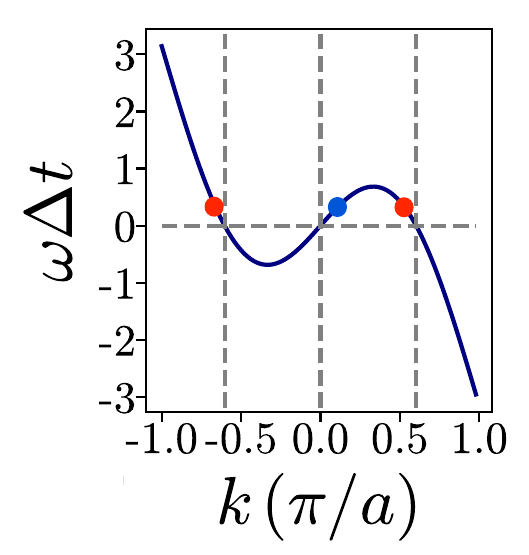}
    \caption{Floquet dispersion on the outside of the black hole horizon. The wave packet in the outside region which is moving towards the horizon during the backwards time evolution consists of plane wave states which lie on the part of the Floquet dispersion near the blue dot. After scattering off the horizon, the reflected wave packet which moves away from the horizon during the backwards time evolution consists of states near the two orange dots.}
    \label{fig:txomegaSupp}
\end{figure}

In Fig.\ \ref{fig:txomegaSupp}, we show the Floquet dispersion relation $\omega(k) = v\sin(ka)/a - v_{Fl}k$ in the outside region, i.e., where $v_{Fl} < v$. The initial positive-energy wave packet at $t=0$ contains eigenstates of the Floquet unitary which lie in the vicinity of the blue dot in Fig.\ \ref{fig:txomegaSupp}. After scattering off the horizon, the reflected wave packet consists of \emph{both positive and negative} momentum states, corresponding to the parts of the Floquet dispersion relation in the vicinity of the two orange dots in Fig.\ \ref{fig:txomegaSupp}. In the Minkowski vacuum, the negative momentum states are occupied, while the positive momentum states are empty. So after scattering off the horizon, the wave packet contains both a positive and negative Minkowski energy part. The negative energy part is responsible for the non zero wave packet occupation number after scattering.

For the calculation of the wave packet occupation numbers in Fig.\ \ref{fig:FLresults}, the width of the wave packets in momentum space was $\sigma = 2(2\pi/L)$. For the calculation of the wave packet correlations across the horizon in Fig.\ \ref{fig:FLresults}, we used $\sigma = 2(2\pi/L)$ for the wave packets behind the horizon, and $\sigma = 2(2\pi/L)\times(v_{in}/v_{out})$ for the wave packets on the outside of the horizon. This relation between the widths of the wave packets is important for the maximal wave packet correlations to be given by $\sqrt{f(\omega)f(-\omega)}$.

Finally, the left-moving wave packet operators $\tilde{W}^\dagger_{x_0,\omega}$ used to obtain the results shown in Fig.\ \ref{fig:WPO} are defined as

\begin{equation}
\tilde{W}^\dagger_{x_0,\omega} = \mathcal{N}\sum_{k = -\pi/a}^{\pi/a} e^{-(k+k^*+\omega/v^*)^2/4\sigma^2} e^{-i k x_0} c^\dagger_k\, ,
\end{equation}
where $v^* = |\partial_k \omega(k)\big|_{k=-k^*}|$ (with $v = v(x\gg x_b)$) is the velocity of the wave packets. The width used in Fig.\ \ref{fig:WPO} was again $\sigma = 2(2\pi/L)$.

\section{Connection between the Corley-Jacobson falling lattice and static horizons in Floquet systems}\label{app:CJ}

In Ref. \cite{CorleyJacobson}, Corley and Jacobson discretized a scalar field on a spatial lattice which is falling into a black hole. As a result of this choice of discretization, the location of the horizon relative to the lattice chances in time. The Corley-Jacobson falling lattice analog of our fermion hopping model used in the main text would therefore correspond to the following time-dependent Hamiltonian:

\begin{equation}
    \hat{H}_{CJ}(t) = \sum_{j=1}^N \frac{\gamma_j(t) + \gamma_{j+1}(t)}{4} (ic^\dagger_{j+1} c_j -i c^\dagger_j c_{j+1})\, ,
\end{equation}
where $\gamma_j(t) = \gamma(ja - v_{Fl}t)$, with $\gamma(x)$ the same continuous function as used in the main text. In the Corley-Jacobson model, the location of the horizon is moving to the right with a velocity $v_{Fl} = a/\Delta t$. Particles in the region to the left of the black hole horizon move slower than $v_{Fl}$, so they can never catch up with the horizon. Particles to the right of the black hole horizon travel faster than $v_{Fl}$.

In the Corley-Jacobson model, the time evolution operator is given by

\begin{equation}
    U_{CJ}(t) = \mathcal{T}e^{-i\int_0^t\mathrm{d}t'\,\hat{H}_{CJ}(t')}\, ,
\end{equation}
where $\mathcal{T}$ is the time-ordering operator. At discrete times $t = n\Delta t$, we can write this time evolution operator as

\begin{equation}
    U_{CJ}(n\Delta t) = \mathcal{T}e^{-i\int_{(n-1)\Delta t}^{n\Delta t}\mathrm{d}t'\,\hat{H}_{CJ}(t')}\cdots \mathcal{T}e^{-i\int_{\Delta t}^{2\Delta t}\mathrm{d}t'\,\hat{H}_{CJ}(t')} \mathcal{T}e^{-i\int_0^{\Delta t}\mathrm{d}t'\,\hat{H}_{CJ}(t')}.
\end{equation}
Using the following property of the Corely-Jacobson Hamiltonian:

\begin{equation}
    \hat{H}_{CJ}(t+\Delta t) = \hat{T}_L^{-1} \hat{H}_{CJ}(t) \hat{T}_L\, ,
\end{equation}
we can rewrite $U_{CJ}(n\Delta t)$ as

\begin{eqnarray}
    U_{CJ}(n\Delta t) & = &  \hat{T}_L^{-n} \left( \hat{T}_L \mathcal{T}e^{-i\int_0^{\Delta t}\mathrm{d}t'\,\hat{H}_{CJ}(t')}\right)^n \\
    & := & \hat{T}_L^{-n} \tilde{U}(\Delta t)^n\\
    & \approx & \hat{T}_L^{-n} U(\Delta t)^n,
\end{eqnarray}
where $U(\Delta t)$ is the Floquet unitary with a static causal horizon that we used in the main text (Eq.\ \eqref{FlU}). So we see that there is a close connection between the Corley-Jacobson approach and the Floquet approach. We have checked that quenching $|\psi_0\rangle$ with either $\tilde{U}(\Delta t)$ or $U(\Delta t)$ produces identical results for the Hawking radiation. The additional unitary $\hat{T}_L^{-n}$ in the Corley-Jacobson time evolution operator simply implements a transformation from the co-moving frame where the horizon is static to the laboratory frame where the horizon is moving to the right.

\section{Details of numerical simulations for the local Hamiltonian model}\label{app:local}

The site-dependent hopping term $\gamma_j = \gamma(ja)$ used in the local Hamiltonian $\hat{H}$ is given by
\begin{equation}
    \gamma_j/\gamma =  1 -2 S(2\hat{\kappa} (j-j_b))-2S(-2\hat{\kappa} (j-j_w)),
    \label{eq:v_localH}
\end{equation}
where $S(x)=\frac{1}{1+e^x}$. $j_b$ ($j_w$) is the location of the black hole (white hole) horizon and $\hat{\kappa}$ determines the slope of the interpolation between $\gamma_j = -\gamma$ and $\gamma_j =\gamma$, such that the surface gravity is given by $\kappa=a\partial_x \gamma(x)|_{x=x_b}=\hat{\kappa}\gamma$.

\begin{figure}
    \centering
    \includegraphics[width=0.8\linewidth]{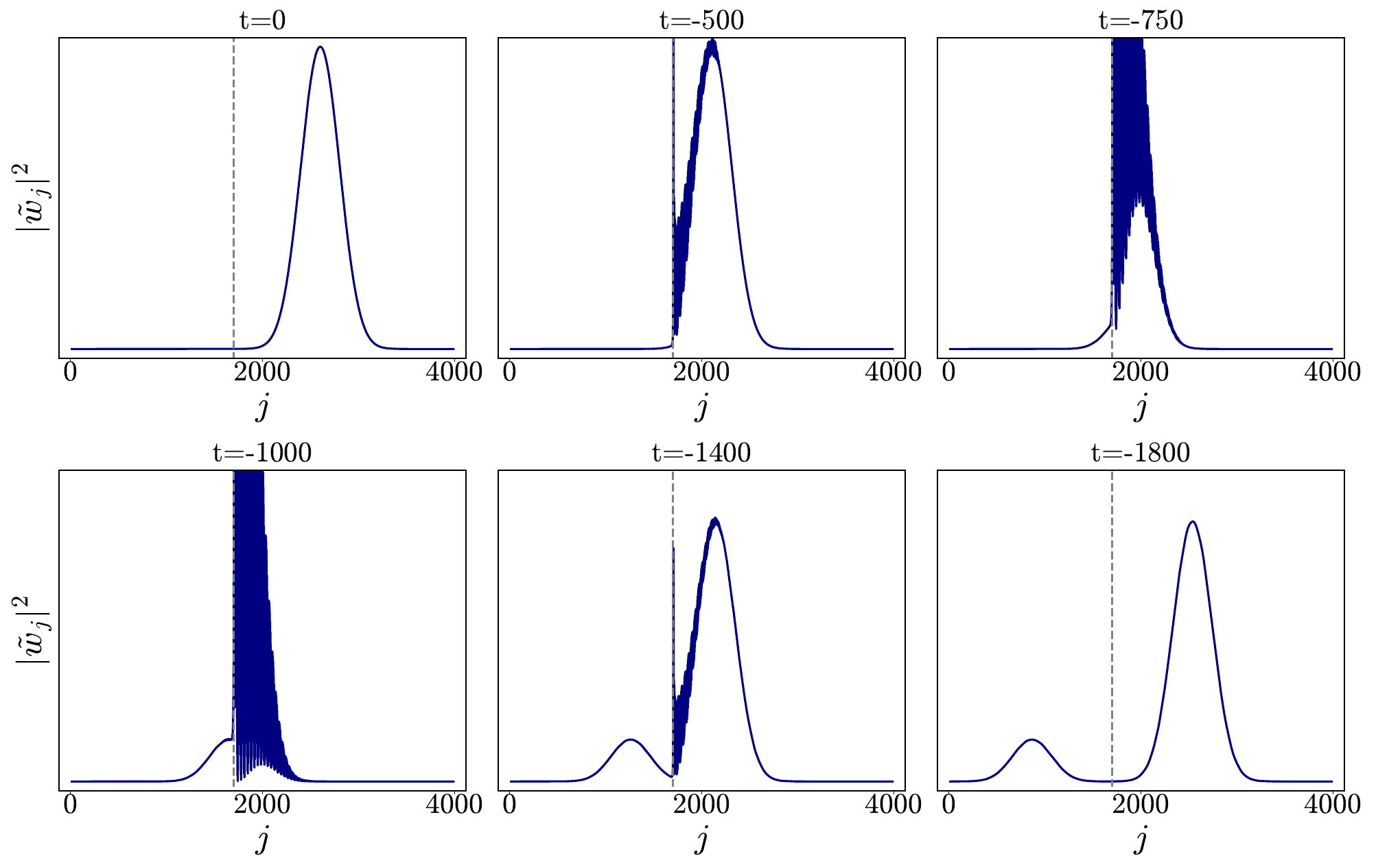}
    \caption{Backwards-in-time evolution of the wave packet $W^\dagger_{x_0,\omega} = \sum_j w_j c^\dagger_j$ with $x_0 = x_b + 900 a$ and $\omega = 0.00298 t$. The width of the wave packet in momentum space is $\sigma = 0.0025a^{-1}$. The system size is $N=4000$, $x_b = 1700 a$, and $x_w = 3900 a$. The hopping $t_j$ in Eq.\ \eqref{eq:v_localH} was used with $\hat{\kappa}=0.1$, together with $\mu=0.5 t$. }
    \label{fig:Backwp_local}
\end{figure}
The wave packet creation operator $W^\dagger_{x_0,\omega}$ is constructed from the momentum modes of the Minkowski Hamiltonian and is defined as
\begin{equation}
    W^\dagger_{x_0,\omega} = \mathcal{N} \sum_{k=-\pi/a}^{k=\pi/a} e^{-(k-k_0)^2/4\sigma^2} e^{-ikx_0} c_k^\dagger\, ,
\end{equation}
with $k_0$ a free parameters that determines the momentum mode around which the wave packet is centered (and consequently whether it is left- or right-moving), and $\sigma$ determines the width of the Gaussian. Here $c_k^\dagger$ is the (discrete) fourier transform of the on-site creation operators i.e. $\tilde{c}^\dagger_k = \frac{1}{\sqrt{N}}\sum_{j=1}^{N}e^{ikaj}c^\dagger_j$. For all numerical calculations, the width $\sigma$ was taken to be $0.0025 a^{-1}$ for $L=4000a$, so that the wave packet is as narrow as possible in momentum space, while still able to fit in the inside region in real space. For $\mu$ we used a value of $0.5 \gamma$.

\begin{figure}
    \centering
    \includegraphics[width=0.85\linewidth]{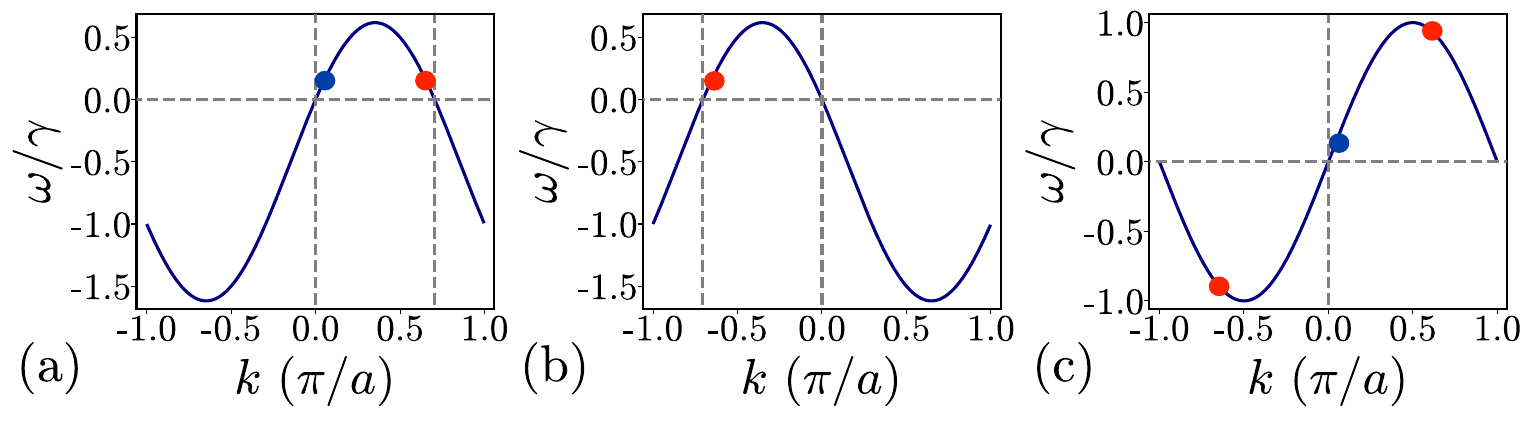}
   \caption{Dispersions for the local model on the outside (a) and inside (b) of the black hole horizon. The right-moving positive-energy wave [blue dot in (a)] packet initially starts outside the horizon and under backwards time evolution scatters of the horizon. The reflected wave packet contains only positive momentum states [orange dot in (a)] while the transmitted part only has negative momentum states [orange dot in (b)]. Their corresponding Minkowski energies are shown on the Minkowski dispersion $\omega_0/t=\sin(k a)$ in (c), with a negative energy for the transmitted part, and positive energies for the incoming and reflected part.}
    \label{fig:Freqslocal_Mink}
\end{figure}

Similar to the Floquet case, the time-dependent wave packet occupation number $N_{x_0,\omega}(t)$ is calculated numerically by transferring the time evolution to the wave packet creation operators $ W^\dagger_{x_0,\omega}$ via $N_{x_0,\omega}(t) =\langle \psi_0 |W^\dagger_{x_0,\omega}(-t) W_{x_0,\omega}(-t)|\psi_0 \rangle$, where $W^\dagger_{x_0,\omega}(-t)=U^\dagger(t) W^\dagger_{x_0,\omega} U(t)$ is the wave packet operator evolved \textit{backwards} in time. An example of a backwards-in-time evolved wave packet is shown in Fig.\ \ref{fig:Backwp_local}. Note that because of the \emph{backwards} time evolution, the right-moving wave packet actually moves to the left. We see that as time proceeds, the wave packet scatters off the boundary between the inside ($\gamma_j<0$) and outside ($\gamma_j>0$) regions, and that only part of it goes through this boundary while the other part is reflected back. Here the reflected wave packet only contains positive momentum modes which lie around the orange dot in Fig.\ \ref{fig:Freqslocal_Mink}(a), while the transmitted part solely consists of negative momentum modes near the orange dot in figure \ref{fig:Freqslocal_Mink}(b). The momentum of the incoming wave packet lies in the vicinity of the blue dot in Fig.\ \ref{fig:Freqslocal_Mink}(a). The corresponding Minkowski energy modes are shown in Fig.\ \ref{fig:Freqslocal_Mink}(c). Since in the Minkowski ground state only negative energy modes are occupied it is clear that it is exactly the transmitted wave packet that gives rise to the non zero occupation number observed. 

\section{Details of the analytical derivation of the transmission coefficient}\label{app:analytical_discr}
\begin{figure}
    \centering
    \includegraphics[width=0.5\linewidth]{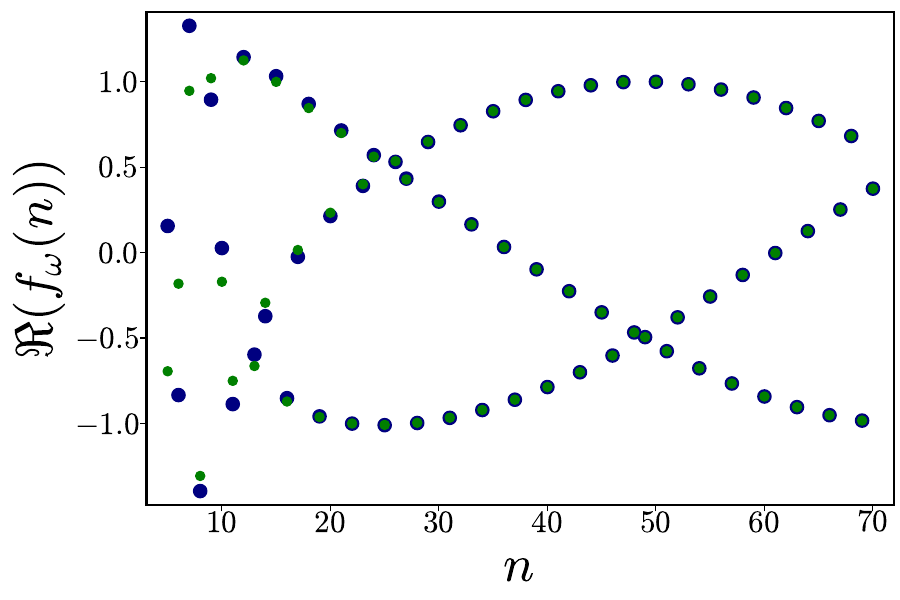}
    \caption{Comparison between the numerical solution of \eqref{eq:app_recurrence} (blue) and the discrete WKB approximation \eqref{eq:ImagWKB} (green) for $n>n_t\approx 3.5$, $\hat{\kappa}=0.1$, $\omega=0.1\gamma$ and $\mu=0.5\gamma$.}
    \label{fig:WKBcomparison}
\end{figure}

The recurrence equation
\begin{equation}
    \tau_n f_\omega(n+1) + \tau_{n-1}f_\omega(n-1) + (\mu+\omega)f_\omega(n) = 0\,,
    \label{eq:app_recurrence}
\end{equation}
can be solved by a discrete version of the well-known Wentzel-Kramers-Brillouin (WKB) approximation (see Refs.\ \cite{Dingle,Braun,Geronimo}). Similar to the continuous WKB approximation one finds that the solutions are either complex or real exponentials, depending on the value of 
\begin{equation}
    B(n)=\frac{\mu+\omega}{2\sqrt{\tau_n \tau_{n-1}}} \approx \frac{\mu+\omega}{\sqrt{\mu^2+\gamma^2\tanh^2 \hat{\kappa}n}}\,.
\end{equation}
Denoting the turning points for which $B(n)=1$ as $\pm n_t$ we have that the discrete WKB ansatz is given by
\begin{equation}\label{eq:discreteWKB}
    f_\omega(n) = \frac{1}{\sqrt[4]{4\tau_n \tau_{n-1}}} 
    \begin{cases}
       C_{1}P^{I,+}_\omega(n)+C_{2}P^{I,-}_\omega(n) ,\quad|n| \gtrsim n_t \\
       \tilde{C}_{1}P^{R,+}_\omega(n)+\tilde{C}_{2}P^{R,-}_\omega(n) ,\quad|n|\lesssim n_t\\
    \end{cases}
\end{equation}
with 
\begin{equation}\label{eq:ImagWKB}
    P^{I,\pm}_\omega(n) = \frac{1}{\sqrt[4]{1-B(n)^2}}\exp\left( \pm i \int^n \acos(-B(y)) \D y\right)\,,
\end{equation}
and
\begin{equation}\label{eq:RealWKB}
    P^{R,\pm}_\omega(n) = \frac{1}{\sqrt[4]{B(n)^2-1}}\exp\left( \pm \int^n \acosh B(y) \D y\right)\,.
\end{equation}
For $|n|\gg n_t$ we have that $f_\omega(n)$ reduces to the plane waves solution \eqref{eq:planewaves} up to some irrelevant factors. A comparison between the numerical solution of \eqref{eq:app_recurrence} and the WKB approximation \eqref{eq:ImagWKB} is shown in Fig.\ \ref{fig:WKBcomparison}. Just like in the continuous WKB approximation the solutions break down near the turning points $\pm n_t$. This prevents the direct matching of the asymptotic solutions for $n\ll -n_t$ and $n\gg n_t$ which determines the transmission coefficient. To circumvent this problem we notice that near the turning points $B(n)\approx 1$ so that we can approximate $\acos ( -B(n) ) \approx \pi+\sqrt{2(1-B(n)}$ and $1-B(n)^2 \approx  2(1-B(n))$. Then
\begin{equation}
    P^{I,\pm}_\omega(|n| \gtrsim n_t) \approx \frac{(-1)^n}{\sqrt{a}}\Psi^{\pm}(x)\,,
\end{equation}
with 
\begin{equation}
    \Psi^{\pm}(x) = \frac{1}{\sqrt{k(x)}}\exp\left( \pm i \int^x k(y) \D y\right)\,,
\end{equation}
where $k(y) =\frac{1}{a} \sqrt{2(1-B(y))}$ has the dimension of 1/length. Here $\Psi^{\pm}(x)$ is readily recognized as a WKB ansatz for a continuous Schrodinger equation with $k(x)=\sqrt{2m(E-V(x))}$. Given that near the turning points $k(x)$ is indeed slowly varying on the lattice scale, this allows us to map the discrete problem around the turning points to a continuous scattering problem with identifications
$m=\frac{\mu+\omega}{\gamma^2 a^2}$, $E=\frac{\gamma^2}{\mu+\omega}$ and a potential
\begin{equation}\label{eq:potential}
    V(x)=\frac{\gamma^2}{\sqrt{\mu^2+\gamma^2\tanh^2 \tilde{\kappa}x}}\,,
\end{equation}
where we have used the shorthand $\tilde{\kappa} = \frac{\hat{\kappa}}{a} = \frac{\kappa}{\gamma a}$. We can now relate the solution for $\Psi^{\pm}(n)$, and thus also $P^{I,\pm}_\omega(n)$, at $n \gtrsim n_t$ and $n \lesssim n_t$ using the transmission coefficient for the continuous problem. Starting from $n \gg n_t$ we have that the discrete solution is given by
\begin{equation}
    f_\omega(n \gg n_t) = A_R e^{ikn} = 
    \frac{1}{\sqrt[4]{4\tau_n \tau_{n-1}}} 
       D_{1}P^{I,+}_\omega(n)\,,
\end{equation}
with $D_1 = \sqrt{2\tau_\infty} e^{i\theta} A_R$ and $\theta$ a phase such that $e^{\pm i\theta} P^{I,\pm}_\omega(n\gg n_t) = e^{\pm ikn}$.
For $n \gtrsim n_t$, $P^{I,+}_\omega(n)$ coincides with $(-1)^n\Psi^+(n) $ so that
\begin{equation}
    f_\omega(n \gtrsim n_t) =
    \frac{1}{\sqrt[4]{4\tau_n \tau_{n-1}}} 
       D_{1}P^{I,+}_\omega(n) \approx \frac{(-1)^n}{a\sqrt[4]{4\tau_n \tau_{n-1}}} 
       D_{1}\Psi^+(n) \,.
\end{equation}
The solution of the continuous scattering problem tells us that 
\begin{equation}
    D_{1}\Psi^+(n \gtrsim n_t) = C_{1}\Psi^+(n \lesssim -n_t)+C_{2}\Psi^-(n \lesssim -n_t)\,,
\end{equation}
with $\frac{|D_1|^2}{|C_1|^2} = T_{\text{cont}}$. Then for $n \lesssim -n_t$
\begin{equation}
    f_\omega(n \lesssim n_t) = \frac{(-1)^n}{a\sqrt[4]{4\tau_n \tau_{n-1}}}  \left(
       C_{1}\Psi^+(n)+C_{2}\Psi^-(n) \right) \approx \frac{1}{\sqrt[4]{4\tau_n \tau_{n-1}}} \left(
       C_{1}P^{I,+}_\omega(n)+C_{2}P^{I,-}_\omega(n) \right)\,,
\end{equation}
where we used that $(-1)^n\Psi^\pm(n)$ and $P_\omega^{I,\pm}$ coincide near the turning point $-n_t$. For $n\ll -n_t$ the discrete solution reduces to 
\begin{equation}
    f_\omega(n\ll -n_t) = \frac{1}{\sqrt{2\tau_\infty}}  \left(
       C_{1}e^{-i\theta}e^{ikn}+C_{2}e^{i\theta}e^{-ikn} \right)\,,
\end{equation}
which gives $B_R = \frac{1}{\sqrt{2\tau_\infty} }C_1e^{-i\theta}$ and $B_L = \frac{1}{\sqrt{2\tau_\infty} }C_2e^{i\theta}$.The transmission coefficient for the discrete problem is then
\begin{equation}
    T_{\text{lat}} = \frac{|A_R|^2}{|B_R|^2} = \frac{|D_1|^2}{|C_1|^2} = T_{\text{cont}}\,.
\end{equation}
For a general rounded potential hill that is approximately parabolic Kemble \cite{Kemble} has shown that the transmission coefficient $T_{\text{cont}}$ is given by
\begin{equation}
    T_{\text{cont}} = 
    \begin{cases}
    \begin{aligned}
        &\frac{1}{1+e^{K}}, \quad E<V_{\text{max}}\\
        &\frac{1}{1+e^{-K'}}, \quad E>V_{\text{max}}\\
    \end{aligned}
    \end{cases}
    \label{eq:Kemble}
\end{equation}
with 
\begin{equation}
    K = 2\int_{x_1}^{x_2} \sqrt{2m(V(x)-E)} \D x \,,
    \label{eq:cont_approx}
\end{equation}
and
$$K' = 2\int_{\hat{x}_1}^{\hat{x}_2} \sqrt{2m(E-V(ix)} \D x \,.$$
Here the integration limits $x_1,x_2$ ($\hat{x}_1,\hat{x}_2$) are the real (complex) zeros of $V(x)-E$, given by $\pm \tilde{\kappa} x_t=\pm  \atanh \frac{1}{\gamma}\sqrt{(\mu+\omega)^2-\mu^2}$ and $\pm \tilde{\kappa} \hat{x}_t=\pm  \atan \frac{1}{\gamma}\sqrt{\mu^2-(\mu+\omega)^2}$. It is easy to show that $K'=-K$ due to the particular form of $V(x)$. Using the identifications made earlier $E<V_{max} $ ($E>V_{max}$) corresponds to $\omega > 0$ ($\omega < 0$) and it follows that
\begin{equation}
    T_{\text{cont}} = \frac{1}{1+e^{K}}, \quad \forall \omega\,.
\end{equation}
For $\omega\ll \mu$ the integration range in $K$ will be small and most of the contribution to the integral will come from $x\approx 0$ where the potential $V(x)$ can safely be approximated as a parabola. In order to get the leading contribution we approximate $x_t$ to the lowest order in $\omega $, $\Tilde{\kappa}\gamma x_t\approx \sqrt{2\mu} \omega^{1/2}$, and $V(x)\approx \gamma^2(\frac{1}{\mu} - \frac{\Tilde{\kappa}^2\gamma^2}{2\mu^3}x^2 )$. Since the integration limits depend on $\omega$ we have that an expansion in terms of $x$ is in effect one in $x\sim\sqrt{\omega}$. Neglecting terms of the order $\mathcal{O}(\omega x^2) = \mathcal{O}(\omega^2)$ we arrive at
\begin{equation}
    K \approx \frac{2\sqrt{2}}{\kappa} \int_{-\sqrt{2\mu\omega}}^{ \sqrt{2\mu\omega}} \sqrt{\frac{\omega}{\mu}-\frac{x^2}{2\mu^2}} \D x = \frac{4\omega}{\kappa} \int_{-1}^{1} \sqrt{1-y^2} \D y = \frac{2\pi  \omega}{\kappa}\,,
    \label{eq:app_leading_order}
\end{equation}
The leading order of the transmission coefficient is thus given by
\begin{equation}
    T_{\text{cont}} = \frac{1}{1+e^{\frac{2\pi  \omega}{\kappa}}} \,.
\end{equation}
Higher order corrections to \eqref{eq:app_leading_order} can be computed by including higher-order terms in the $\omega$ expansions for $x_t$ and $V(x)$. For example up to order $\omega^3$ we have
\begin{equation}\label{eq:next_to_leading}
    K \approx \frac{2\pi \omega}{\kappa}\left(1+\left( \frac{\gamma}{16\mu}+\frac{\mu}{2\gamma}\right) \frac{\omega}{\gamma} \right)\,,
\end{equation}
We observe that the above correction does not agree with the transmission coefficient calculated from the numerical solution of \eqref{eq:app_recurrence}. Inspired by the form of the discrete solutions for $|n|<|n_t|$, and the result of Kemble \cite{Kemble} for the continuous case, we propose that  
\begin{equation}
    K = \frac{2}{a}\int_{x_1}^{x_2} \acosh B(x) \D x \,,
    \label{eq:acosh_approx}
\end{equation}
i.e.\ replacing $\sqrt{2(1-B(x))}$ by $\acosh B(x)$, will provide a more accurate result for the observed transmission coefficient compared to the continuum approximation \eqref{eq:cont_approx}. Using differentiation under the integral sign we find for \eqref{eq:acosh_approx} that
\begin{equation}
\frac{\D K}{\D \omega} = \frac{2\pi}{\hat{\kappa}} \frac{1}{\sqrt{\gamma^2+\mu^2-(\mu+\omega)^2}}\,,
\end{equation}
which already indicates that $K \approx \frac{2\pi}{\kappa} \omega + \mathcal{O}(\omega^2)$, reproducing the leading order term in \eqref{eq:app_leading_order}. Integrating the above expression gives
\begin{equation}\label{eq:app_expfactor}
    K = \frac{2\pi}{\hat{\kappa}}\left (\atan \frac{\mu+\omega}{\sqrt{\gamma^2+\mu^2-(\mu+\omega)^2}} -\atan \frac{\mu}{\gamma}\right)\,, 
\end{equation}
where we set the integration constant to $-\atan \mu/\gamma$ since the original integral for $K$ is trivially zero for $\omega=0$. For small $\omega$ Eq.\ \eqref{eq:app_expfactor} becomes
\begin{equation}
    K = \frac{2\pi}{\hat{\kappa}} \left(\frac{\omega}{\gamma} + \frac{\mu}{2\gamma} \frac{\omega^2}{\gamma^2}\right) = \frac{2\pi \omega}{\kappa} \left(1 + \frac{\mu}{2\gamma} \frac{\omega}{\gamma} \right)\,,
\end{equation}
highlighting the difference with \eqref{eq:next_to_leading}.

\end{appendix}

\end{document}